\documentclass{emulateapj}

\usepackage{natbib}

\newcommand{\be}{\begin{equation}}
\newcommand{\ee}{\end{equation}}

\newcommand{\ifm}[1]{\relax\ifmmode#1\else$\mathsurround=0pt #1$\fi}
\newcommand{\kms}{\ifmmode\,{\rm km}\,{\rm s}^{-1}\else km$\,$s$^{-1}$\fi}
\newcommand{\kpc}{\ifmmode\,{\rm kpc}\else kpc\fi}
\newcommand{\Mpc}{\ifmmode\,{\rm Mpc}\else kpc\fi}

\newcommand{\ltsima}{$\; \buildrel < \over \sim \;$}
\newcommand{\lsim}{\lower.5ex\hbox{\ltsima}}
\newcommand{\gtsima}{$\; \buildrel > \over \sim \;$}
\newcommand{\gsim}{\lower.5ex\hbox{\gtsima}}

\long\def\symbolfootnote[#1]#2{\begingroup\def\thefootnote{\fnsymbol{footnote}}\footnote[#1]{#2}\endgroup}

\def\apjl{\textit{Astrophys. J., Lett.}}        
\def\apjs{\textit{Astrophys. J., Suppl. Ser.}}  
\def\aap{\textit{Astron. Astrophys.}}          

\shorttitle{Tidal debris in elliptical galaxies}
\shortauthors{Feldmann, Mayer, Carollo}
\journalinfo{The Astrophysical Journal, 684:1062-1074, 2008 September 10}
\submitted{}

\begin{document}

\title{Tidal debris in elliptical galaxies as tracers of mergers with disks}


\author{R. Feldmann\altaffilmark{1}, L. Mayer\altaffilmark{2} and C. M. Carollo\altaffilmark{1} }


\altaffiltext{1}{Department of Physics, Institute of Astronomy, 
                 ETH Z\"urich, 
		 Wolfgang-Pauli-Strasse 16, 8093 Z\"urich, Switzerland; feldmann@phys.ethz.ch}
\altaffiltext{2}{Department of Physics, Institute of Theoretical Physics, 
                 University of Zurich, 
                 Winterthurerstrasse 190, 8057 Z\"urich, Switzerland}	 


\begin{abstract}
We use a set of high-resolution $N$-body simulations of binary galaxy mergers to show that the morphologies of the tidal features  that are seen around a large fraction of nearby, massive ellipticals  in the field, cannot be reproduced by equal-mass dissipationless mergers; rather, they are well explained by the accretion of disk-dominated galaxies. In particular, the arm- and looplike morphologies of the observed tidal debris can only be  produced by the kinematically cold material of the disk components of the accreted galaxies. The tidal features that arise from such ``cold-accretion'' events onto a massive elliptical are visible for significantly longer timescales than the features produced by elliptical-elliptical mergers (about 1-2 Gyr vs. a few hundred million years). Mass ratios of the order of $1:10$ between the accreting elliptical and the accreted disk are sufficient to match the brightness of the observed debris. Furthermore,  stellar population synthesis models and simple order-of-magnitude calculations  indicate that the colors of the tidal features generated in such minor cold-accretion events are relatively red, in agreement with the observations.  The minor cold-accretion events  that explain the presence, brightness, and structural and color properties of the tidal debris cause only a modest mass and luminosity increase in the accreting massive elliptical.  These results, coupled with the relative statistical frequencies of disk- and bulge-dominated galaxies in the field, suggest that massive ellipticals assemble most of their mass well before their tidal debris forms through the accretion of relatively little, kinematically cold material rather than in very recent, dissipationless major mergers.
\end{abstract}



\section{Introduction}

Bright elliptical galaxies are the most massive stellar systems in the Universe and host a 
large fraction of the stellar mass that has been produced since the Big Bang~\citep{1998ApJ...503..518F,2006ARAaA..44..141R}. 
The boxy isophotes and slow internal rotation of the most massive ellipticals argue for dissipationless mergers playing a role in the last stage of their assembly history~\citep{2006ApJ...636L..81N,2005MNRAS.359.1379K}. 
Semianalytic models indicate that major mergers that lead to massive ellipticals are typically not disk-disk mergers but rather elliptical-elliptical or mixed mergers~\citep{2003ApJ...597L.117K}. Another scenario, as suggested by cosmological SPH simulations~\citep{2007ApJ...658..710N}, is the  early build-up of ellipticals at high redshift with only minor growth in mass due to small accretion events below $z<1$.
The red, stellar tidal arms and loops in the outskirts of many nearby bright ellipticals are widely believed to be fast-dissolving transient beacons of \emph{recent} dissipationless mergers between smaller, equal-mass ellipticals~\citep{2005AJ....130.2647V}.\footnote{The merger rates derived in~\cite{2005AJ....130.2647V} are calculated from a close galaxy pair analysis and hence should not suffer from a misidentification of the origin of the tidal features.} Such mergers are often called ``dry'' to highlight the absence of gaseous material (as opposed to ``wet'', or gas-rich, mergers). Numerical simulations of elliptical-elliptical mergers indicate that tidal debris disappears within a few hundred million years~\citep{1995AaA...297...37C,2005AJ....130.2647V,2006ApJ...640..241B}. This would imply a substantial increase in the massive elliptical galaxy population in the last few billion years~\citep{2004ApJ...608..752B}. This picture faces, however, several observational challenges~\citep{2006AaA...453L..29C,2007ApJS..172..494S}.

Other studies~\citep{2006ApJ...648..969K} had hinted at the gradual disruption of a tiny satellite galaxy by a massive elliptical as a source for red tidal features. Yet, these disruption events yield tidal debris too faint to match the typical surface brightnesses of the observed tidal features~\citep{2005AJ....130.2647V}. Early numerical experiments~\citep{1988ApJ...331..682H} showed that sharp tidal arms and loops could arise from the tidal disruption of dynamically cold stellar disks. These experiments however adopted a spherical model for the elliptical galaxy and lacked the proper resolution to robustly establish the morphology and lifetime of the tidal streams. Moreover, those early studies were carried out
before the current cosmological framework, the cold dark matter model with a cosmological constant, was established.

We use a set of 13 high-resolution $N$-body simulations of galaxy-galaxy mergers between a massive elliptical galaxy and an elliptical or disk galaxy companion to investigate the origin of tidal features observed in bright ellipticals. 
The companion mass ranges between 10\% and 100\% of the mass of the main elliptical (Table~\ref{tab:modParam}). The numerical galaxy models are constructed to properly match the observed structural properties of both elliptical~\citep{1992ApJ...399..462B} and disk galaxies~\citep{1998MNRAS.295..319M}; in particular, the main massive elliptical has a stellar mass of $2.7\times{}10^{11}$ $M_\odot$, an effective radius of 9 kpc, and a stellar mass-to-light ratio of 4 in the $R$-band. These parameters ensure that the model galaxy lies on the observed fundamental plane of ellipticals~\citep{1987ApJ...313...59D,1987ApJ...313...42D}. The stellar component of the galaxies is embedded in a cold dark matter halo with properties derived from cosmological simulations~\citep{1997ApJ...490..493N}.  The initial orbits in the mergers are set to be either parabolic, as expected in low-density environments, or eccentric, with an apocenter-to-pericenter ratio of 6:1, as is typically seen in cosmological simulations in denser environments, i.e., in galaxy groups and in the cores of galaxy clusters~\citep{1998MNRAS.300..146G}.  

\begin{table*}
\begin{center}
\caption{Parameters of the Disk Galaxy Models\label{tab:modParam}}
\begin{tabular}{lcccccccccccccc}
\tableline
\tableline
& & & & & & & & & & & & & &\\[-0.20cm]
& $r_{200}$ & & & & & & & & & & & $h_\textrm{DM}$ & $h_{D}$ & $h_{B}$ \\ 
Description & (kpc) & $c$ & $\lambda$ & $m_D$ & $R_d$ & $z_0/R_d$ & $m_B$ & $r_b/R_d$ & $N_\textrm{DM}$ & $N_{D}$ & $N_{B}$ & (kpc) & (kpc) & (kpc) \\ 
& & & & & & & & & & & & & &\\[-0.20cm] \tableline & & & & & & & & & & & & & &\\[-0.12cm]
1:4 disk/bulge & 206.3  & 12 & 0.031 & 0.04  & 2.59 & 0.1 & 0.008 & 0.2 & $1\times{}10^6$ & $1\times{}10^5$ & $1\times{}10^5$ & 0.13 & 0.13 & 0.13 \\
1:4 disk   & 273        & 8  & 0.0398 & 0.03 & 5.64 & 0.15 & 0   & --- & $2.8\times{}10^6$ & $2\times{}10^5$ & 0              & 0.3  & 0.2 & --- \\
1:10 disk/bulge & 171.4 & 12 & 0.031 & 0.04  & 1.84 & 0.1 & 0.008 & 0.2 & $1\times{}10^6$ & $1\times{}10^5$ & $1\times{}10^5$ & 0.13 & 0.13 & 0.13 \\
1:10 disk  & 200        & 8  & 0.0398 & 0.03 & 4.14 & 0.15 & 0   & --- & $2.8\times{}10^6$ & $2\times{}10^5$ & 0              & 0.3  & 0.2 & --- \\ 
& & & & & & & & & & & & & &\\[-0.20cm]\tableline
\end{tabular}
\tablecomments{The first column describes the model type and its (stellar) mass ratio
with respect to the mass of the massive elliptical used in the merger set up. The other columns denote the virial
radius, the concentration, the halo spin, the disk mass fraction, the disk scale length, the disk height in units of
disk scale length, the bulge mass fraction, the bulge scale length in units of disk scale length, the number of dark,
disk, and bulge particles, and their gravitational softenings.
We assume that disk material and dark matter have the same specific angular momentum, i.e. $j_D=m_D$. The Hubble 
parameter is $H=70$ km s${^-1}$ Mpc$^{-1}$.}
\end{center}
\end{table*}

This paper is organized as follows: In section \ref{sect:initialCond} we describe the set-up of the simulated mergers. In section \ref{sect:analysis} we define the tidal parameter and the merger time, which are of relevance in the further discussion. In section \ref{sect:tidalStrength} we discuss the strength and lifetime of the tidal features, their morphology in section \ref{sect:tidalMorph}, and their expected $B-R$ color evolution in section \ref{sect:tidalColor}. In section \ref{sect:fraction} 
we estimate the expected relative frequency of elliptical-elliptical and elliptical-disk mergers based on statistical studies of galaxies in the field. We discuss our results and conclude in section \ref{sect:discussion}.

\section{Initial conditions}
\label{sect:initialCond}

Our simulations have been carried out on the high-\-per\-formance 
Linux cluster Gonzales
using the parallel $N$-body code PKDGRAV~\citep{2001PhDT........21S}, the gravity module of the TreeSPH-code GASOLINE~\citep{2004NewA....9..137W}. 
The galaxy models used in the collisions 
consist of a stellar system embedded in a
dark matter halo with an NFW~\citep{1997ApJ...490..493N} profile.
The stellar system is either an exponential disk, a disk with a bulge with 
Hernquist~\citep{1990ApJ...356..359H} profile, or a stellar model of an elliptical galaxy constructed as described below.
The properties of the disks are derived analytically~\citep{1998MNRAS.295..319M} from the dark matter halo properties.
$N$-body realizations of the specified galaxy models are initialized with appropriate velocities and positions of
the various particles to obtain an equilibrium model using a standard procedure~\citep{1993ApJS...86..389H}
widely used in the literature.
This approach leads to a compound galaxy system that is nearly in equilibrium right from the beginning. 
To minimize the spurious effects due to small departures from exact equilibrium
we evolved our models for several Gyr in isolation
before using them in the merger simulations.
Each of our disk models contains 1-3 million dark matter and star particles and 
the typical force softening is 100-300 pc. We explicitly checked that with the chosen mass and force resolution
our simulations are not significantly affected by artificial two-body relaxation~\citep{1996ApJ...457..455M}, which could cause  
spurious heating and broadening of cold, sharp tidal features. The structural properties of our disk models are 
enumerated in Table~\ref{tab:modParam}.

\begin{figure*}[htbp]
\begin{center}
\begin{tabular}{cc}
\includegraphics[width=65mm, bb=138 228 445 550]{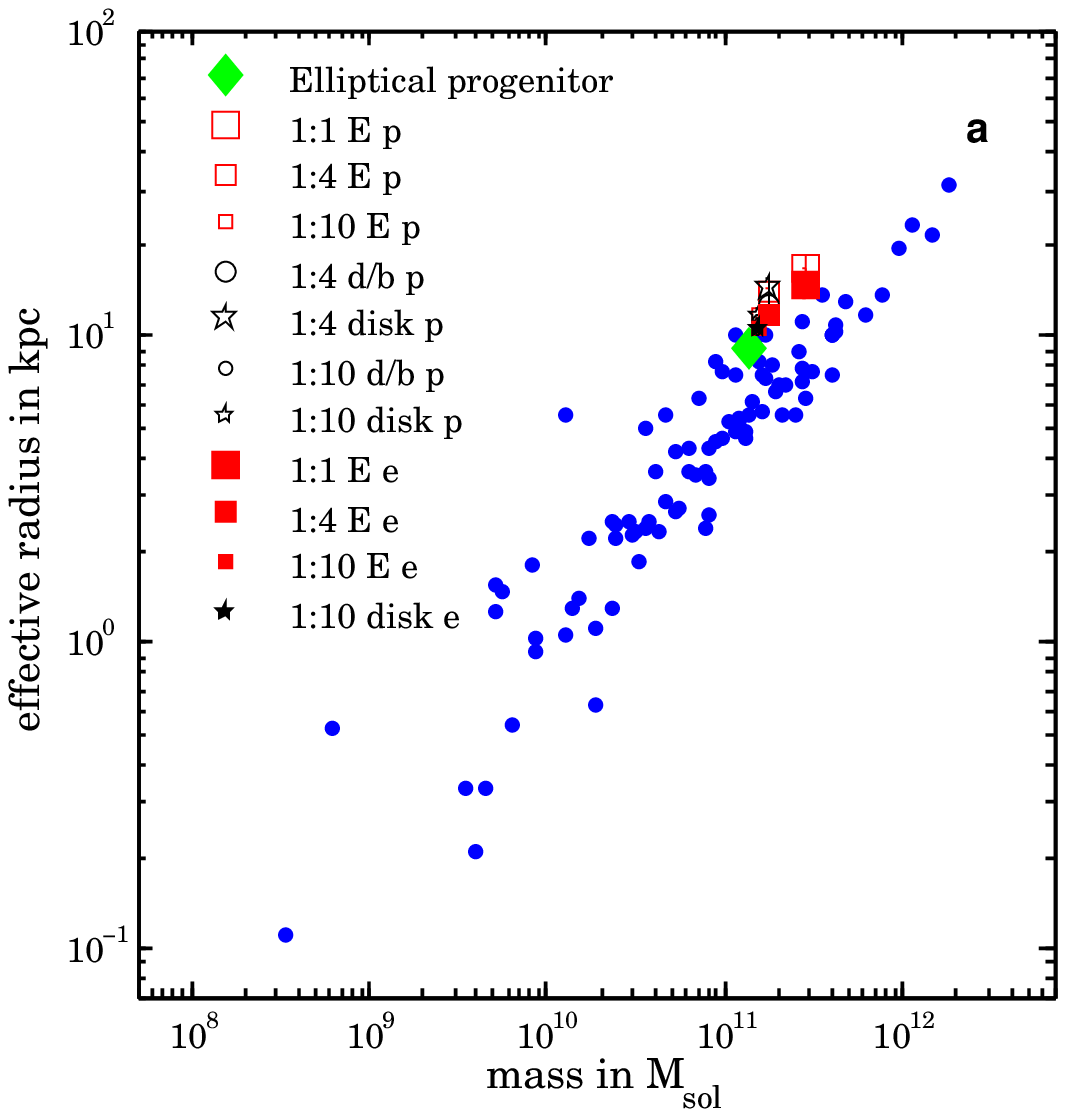} & \includegraphics[width=65mm, bb=138 228 445 550]{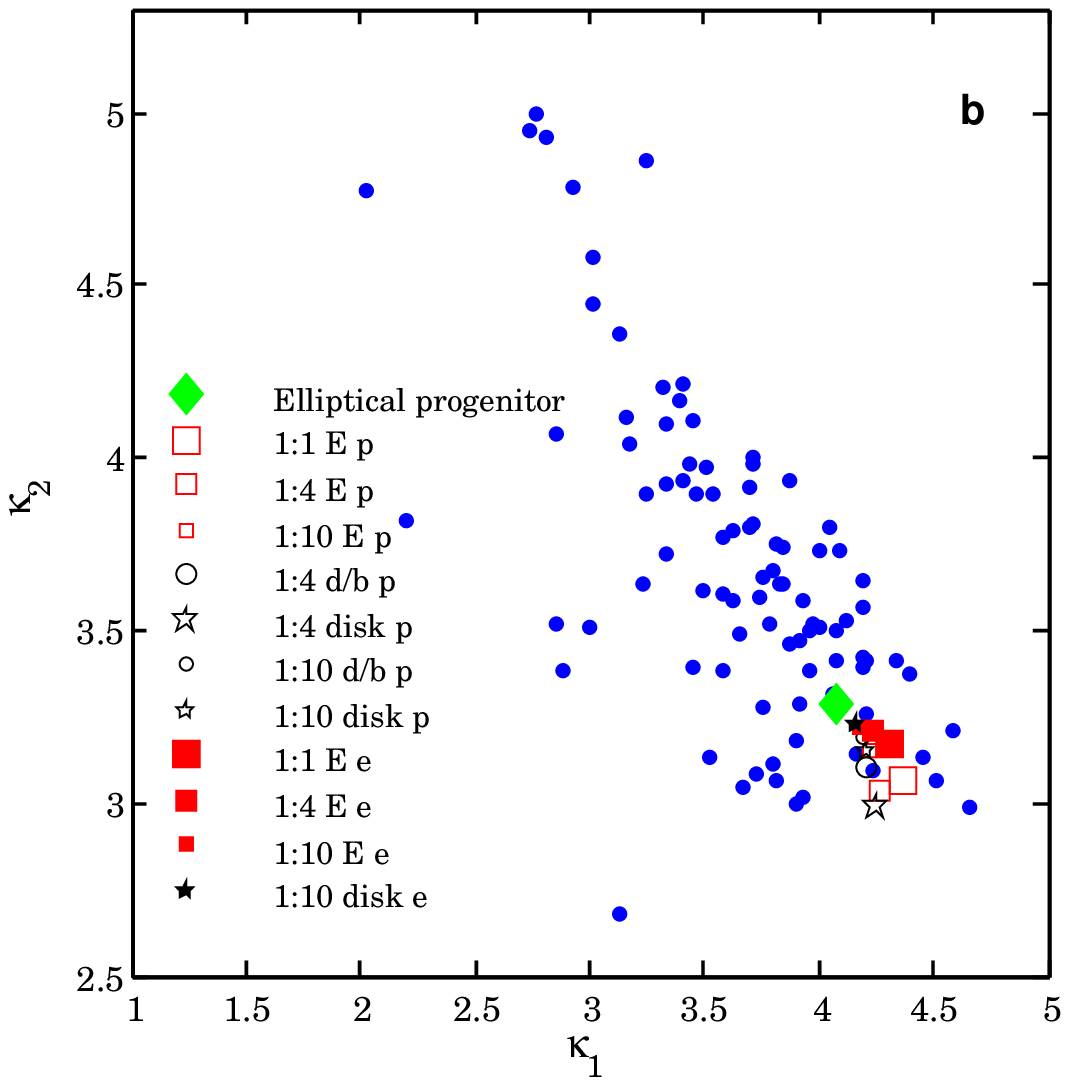} \\
\includegraphics[width=65mm, bb=138 228 445 550]{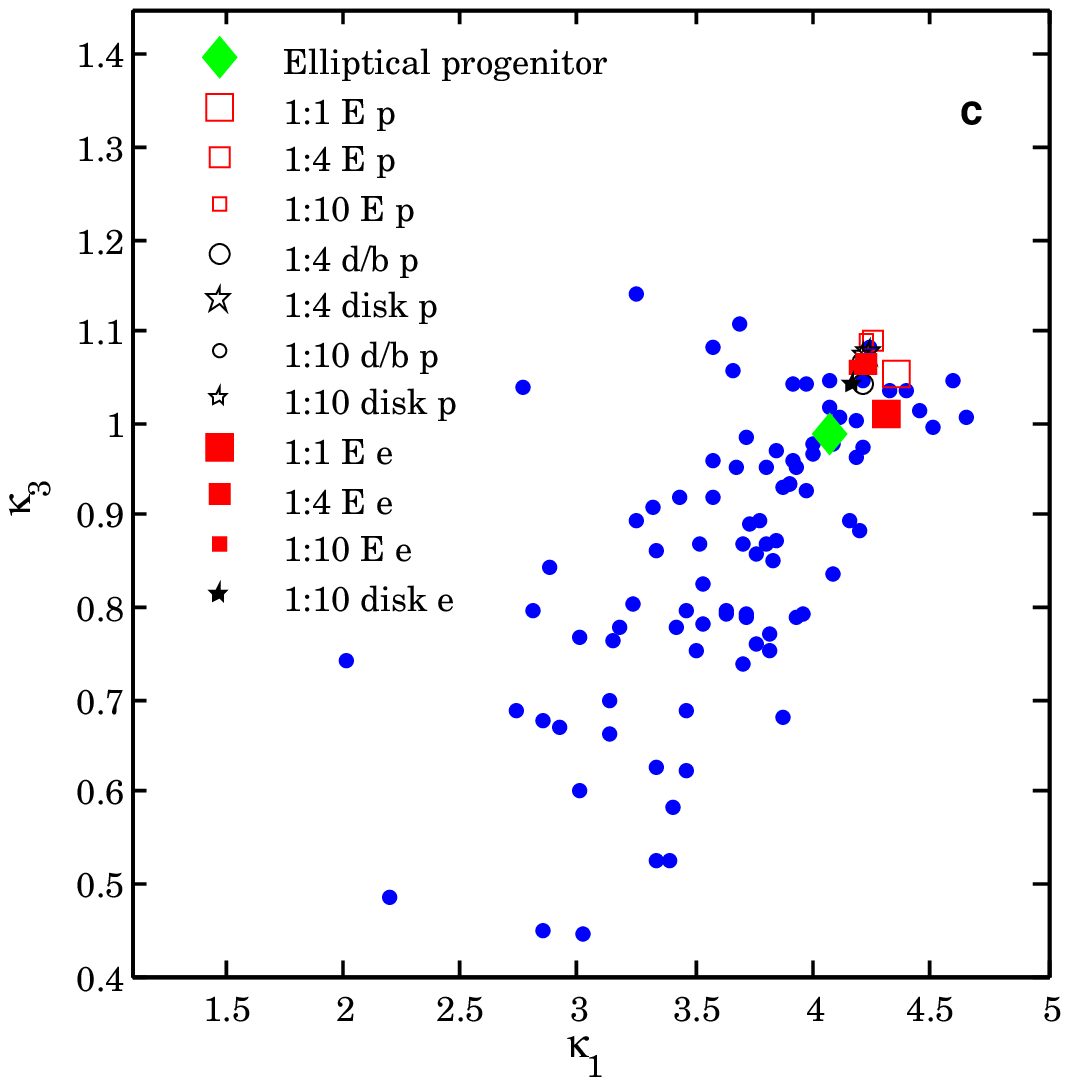} & \includegraphics[width=65mm, bb=138 228 445 550]{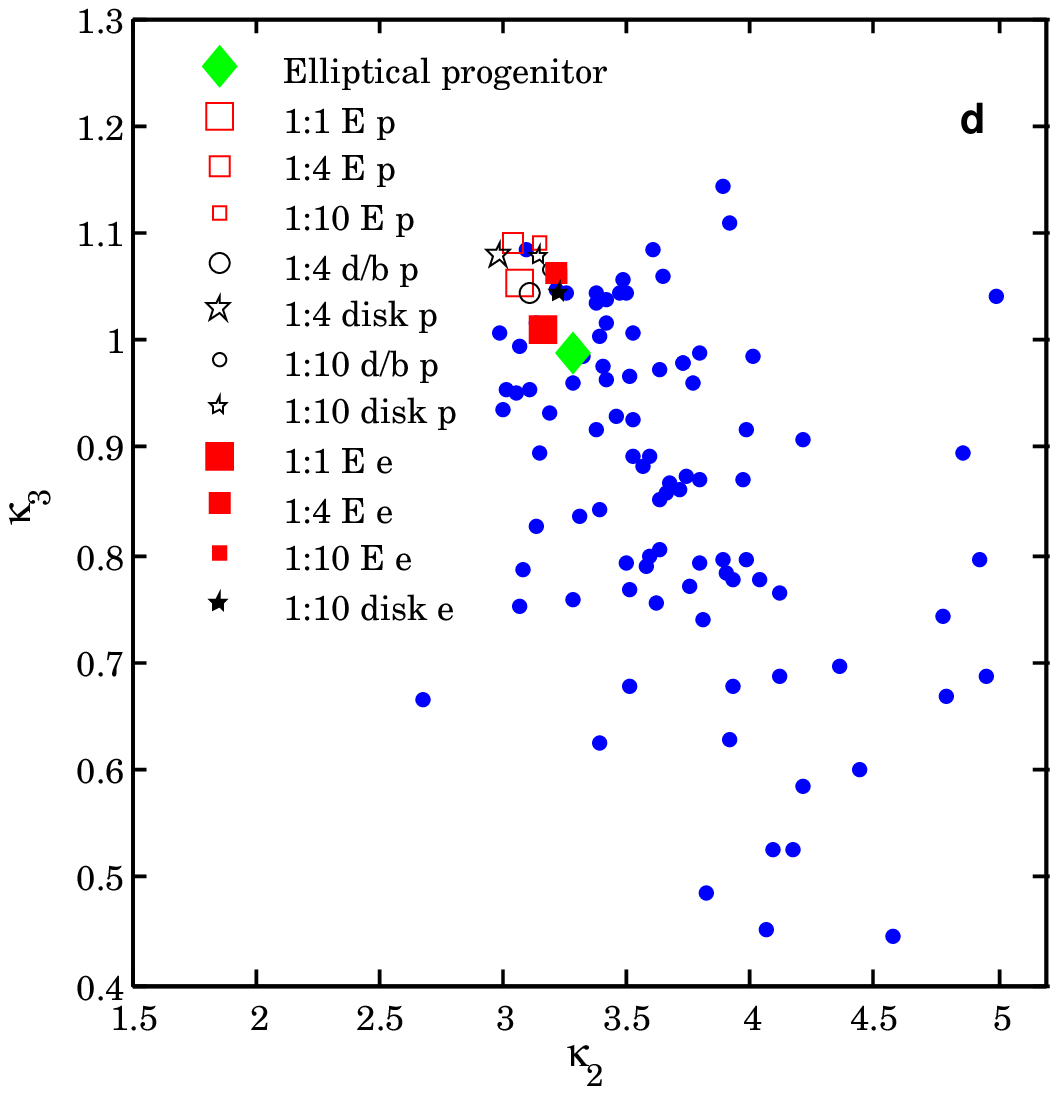} \\
\end{tabular}
\caption{Scaling properties of our merger remnants.
Blue circles show the properties of observed massive elliptical galaxies~\citep{1992ApJ...399..462B}. The position of our massive elliptical 
galaxy model is indicated with the green diamond. Other symbols denote our simulated merger remnants.
The mass ratio (1:1, 1:4, 1:10), the type of the satellite progenitor (elliptical [E], disk+bulge [d/b]
or pure disk [disk]) and whether the orbit is parabolic (p) or eccentric (e) is indicated for each remnant.
($a$) Our massive elliptical host galaxy and its remnants obey 
roughly the typical mass-radius relation of massive elliptical galaxies~\citep{1992ApJ...399..462B}.
($b$-$d$) The merger remnants lie well within the fundamental plane defined by the
$\kappa$-parameters~\citep{1992ApJ...399..462B} $\kappa_1$, $\kappa_2$ and $\kappa_3$. They span an orthogonal coordinate system in the
three-space of the logarithm of the central velocity dispersion, the logarithm of the effective radius, and the mean surface brightness within the effective radius. Here $\kappa_1$, $\kappa_2$, and $\kappa_3$ are
approximately proportional to the logarithm of the stellar mass, the mean surface brightness, and 
the logarithm of the mass-to-light ratio, respectively.}
\end{center}
\label{fig:ScalingProp}
\end{figure*}

Rather than attempting an ``ab initio'' approach in a fully cosmological framework, our
massive elliptical galaxy is built simply by
merging two equal-mass, spherical stellar systems embedded in dark matter NFW halos.  
This is the same procedure that has been successfully applied to build models of
triaxial dark matter halos starting from spherical halos~\citep{2004MNRAS.354..522M} 
and results in a triaxial, boxy stellar system with a small degree of rotation.
The generated model (and its subsequent merger remnants) satisfies the main observational properties 
of luminous ellipticals, including the location on the fundamental plane (Fig.~\ref{fig:ScalingProp}).

The elliptical galaxy model has a stellar mass of $2.7\times{}10^{11}$ $M_\odot$ and an effective radius of 9 kpc and we set the stellar mass-to-light ratio to 4 in the $R$-band, these numbers 
being typical for massive elliptical galaxies~\citep{1992ApJ...399..462B}. The galaxy is resolved with 2 million dark matter and 
2 million star particles, and the force softening is 210 pc for 
dark matter particles and 60 pc for star particles.
Smaller mass elliptical galaxies are obtained by rescaling particle masses, positions and velocities and softening lengths 
of the massive elliptical based on the general scaling relations for galaxies formed in a cold dark matter model~\citep{1998MNRAS.295..319M,2002MNRAS.336..119M}.

A merger between two initial galaxy models is prepared by regarding the galaxies as mass points with masses of $\mu$
times the galaxy mass. The rescaling factor takes into account that in case of close initial positions the galaxies 
are penetrating each other's dark matter halo. The two-body problem is solved for the given 
orbital parameters $d_{init}$, $e$, and $d_{peri}$. Here $d_{init}$ refers to the initial distance 
between the centers of masses of the galaxies, while $d_{peri}$ is the distance at closest 
approach, the so-called pericenter distance. Here $e$ specifies the eccentricity of the orbit.
The galaxy models are then placed at the location of the point masses with their velocities, and 
additional configuration parameters, e.g. the inclination of the disk with respect to 
the orbital plane, are assigned.

We consider two main classes of orbital configurations. The first resembles the nearly-parabolic encounter of two 
galaxies as expected from cosmological simulations to occur mainly in the field~\citep{2006A&A...445..403K}. Our second class of merger set-ups
deals with lower energy, more circular orbits such as those expected for galaxies in virialized groups and 
clusters of galaxies~\citep{1998MNRAS.300..146G}. The apocenter-pericenter ratio of about $6:1$ is 
typical for the latter category of systems. The merger parameters are itemized in Table~\ref{tab:mergParam}.

\begin{table}
\begin{center}
\caption{The Orbital Parameters of the Various Merger Setups\label{tab:mergParam}}
\begin{tabular}{lcrccc}
\tableline
\tableline
& & & & &\\[-0.20cm]
& & $r_\textrm{init}$ & $r_\textrm{peri}$ & & \\ 
Description & Orbit & (kpc) & (kpc) & $e$ & $\mu$ \\
& & & & &\\[-0.20cm] \tableline & & & & &\\[-0.12cm]
(1:1) E + E & p & 700 & 50 & 0.95 & 1\\
(1:4) E + E & p & 518 & 18 & 0.99 & 1\\
(1:10) E + E & p & 518 & 18 & 0.99  & 1\\ 
& & & & &\\[-0.20cm] \tableline & & & & &\\[-0.20cm]
(1:4) E + disk/bulge, &  &  &  &  & \\
\phantom{(1:4) }rperi 9 kpc & p & 518 & 9 & 0.99 &  1\\
(1:4) E + disk/bulge & p & 518 & 18 & 0.99 &  1\\
(1:4) E + disk/bulge,& & & & & \\
\phantom{(1:4) }rperi 36 kpc & p & 518 & 36 & 0.99 &  1\\
(1:4) E + disk & p & 518 & 18 & 0.99 & 1 \\
(1:10) E + disk/bulge & p & 518 & 18 & 0.99 &  1\\
(1:10) E + disk & p & 518 & 18 & 0.99 &  1\\
& & & & &\\[-0.20cm] \tableline & & & & &\\[-0.20cm]
(1:1) E + E & e & 250 & 35.7 & 0.75  &  1\\
(1:4) E + E & e & 150 & 15   & 0.818 &  0.4\\
(1:10) E + E & e & 150 & 15 & 0.818 &  0.4\\ 
& & & & &\\[-0.20cm] \tableline & & & & &\\[-0.20cm]
(1:10) E + disk & e & 150 & 15 & 0.818 & 0.4\\ 
& & & & &\\[-0.20cm] \tableline
\end{tabular}
\tablecomments{The columns contain: the identification of the respective merger, whether the orbit is parabolic (p) or eccentric (e), the initial distance, the Keplerian pericentric distance, the Keplerian eccentricity, and the mass reduction factor $\mu$ of the major progenitor (see text). The first simulation is taken from~\citet{2005ApJ...623L..67K}. The inclination angles between orbital and spin plane for the major and minor participant are chosen randomly in our simulations but fixed for a given orbital type (parabolic or eccentric) with the constraint that coplanar or polar configurations are avoided. The pericentric arguments are also chosen randomly.}
\end{center}
\end{table}

\section{Analysis}
\label{sect:analysis}
We define the strength of tidal features by means of a tidal parameter $\theta$ 
based on two-dimensional projections of the mass density. We first project the stellar component 
of our $N$-body system onto a two-dimensional pixelized grid assuming a constant mass-to-light ratio of 4 in the 
$R$-band for all stellar particles. This is done with an adaptive cloud-in-cell algorithm,
where the kernel scale is chosen according to the local density. The images have a resolution of
0.5 kpc pixel$^{-1}$, or a scale of 0\farcs{}27 pixel$^{-1}$ if we assume that our mergers take place at 
redshift $z\sim{}0.1$. Each image is first corrected for $(1+z)^4$ redshift dimming and
$K$-correction (using ZEBRA; \citealt{2006MNRAS.372..565F}), smoothed with a 
Gaussian kernel with FWHM of 1\farcs{}1, in order to mimic the seeing conditions, and finally 
degraded with a background noise of 29.5 mag arcsec$^{-2}$. This generates our ``real'' galaxy image $G$.
We then mask all pixels that have a flux lower than the background noise. Next we apply the 
reverse unsharp masking technique and mask all pixels that deviate by more than a factor of 2 after 
smoothing the image with a Gaussian kernel. Pixels neighboring a masked pixel are also masked.
In the next step we fit elliptical isophotes to our image using the well-known \texttt{ellipse} task of IRAF.
We fix the center at the position of highest stellar density and allow the ellipticity and the 
position angle to vary freely. This results in the model image $M$. If the two merger constituents have 
not merged we apply the described technique to each object individually and add the model images to create $M$. 
A distortion image $F$ is created by first dividing $G$ by $M$ pixelwise and then convolving it with a quadratic median 
filter of $5\times{}5$ pixels to reduce pixel-to-pixel variation. The tidal parameter $\theta$ is now defined as 
$\theta=\langle\vert{}F-1\vert{}\rangle{}$. It measures the (absolute) deviation 
of the fractional residuals over the nonmasked pixels of the image.

As an operational definition of weak and strong tidal features we use values of $\theta=0.10$ and $0.15$ as 
our ``weak'' and ``strong'' tidal feature limits. The former limit corresponds to weak, barely visibly tidal features and the latter to 
strong, easily identifiable tidal features. For comparison, we measured a value of $\theta=0.085\pm{}0.002$ 
for our massive elliptical galaxy model before the interaction begins.
The lifetime of weak/strong tidal features is
understood as the characteristic time during which the median of $\theta$ over the various projections is 
larger than the weak/strong feature limit. The result of this analysis is shown in Table~\ref{tab:tidalTimes}.

\begin{table*}[htbp]
\begin{center}
\caption{The Lifetimes in Strong, Intermediate and Weak Tidal Features for the Different Simulations\label{tab:tidalTimes}}
\begin{tabular}{lcccc}
\tableline
\tableline
& & & & \\[-0.20cm]
 &  & Weak ($\theta>0.1$)& Intermediate ($\theta>0.13$)& Strong ($\theta>0.15$)\\ 
Description & Orbit &(Myr)&(Myr)&(Myr)\\ 
& & & & \\[-0.20cm] \tableline & & & & \\[-0.12cm]
1:1 E + E & p & $1359_{-5.74}^{+229}$ & $74.76_{-24}^{+77.2}$ & $25.01_{-24.8}^{+10.9}$ \\
1:4 E + E & p & $1339_{-32.2}^{+114}$ & $485.3_{-25.2}^{+64.1}$ & $216.2_{-8.32}^{+29.8}$ \\
1:10 E + E & p & $0_{-0}^{+4.7}$ & $0_{-0}^{+0}$ & $0_{-0}^{+0}$ \\ 
& & & & \\[-0.20cm] \tableline & & & &\\[-0.20cm]
1:4 E + disk/bulge, rperi 9 kpc & p & $3150_{-43.9}^{+75.9}$ & $1472_{-36.8}^{+82.4}$ & $1161_{-30.2}^{+38.3}$ \\
1:4 E + disk/bulge              & p & $2692_{-85.7}^{+83.2}$ & $2109_{-78.1}^{+62.9}$ & $1508_{-33.6}^{+66.9}$ \\
1:4 E + disk/bulge, rperi 36 kpc & p & $>1980$ & $1636_{-33}^{+90.3}$ & $1287_{-37.4}^{+65.8}$ \\
1:4 E + disk & p & $3196_{-47.1}^{+39.9}$ & $1906_{-21}^{+121}$ & $1537_{-37.8}^{+43.7}$ \\
1:10 E + disk/bulge & p & $413.2_{-95.3}^{+5.56}$ & $0_{-0}^{+0}$ & $0_{-0}^{+0}$ \\ 
1:10 E + disk & p & $2028_{-76.3}^{+63.1}$ & $1007_{-17.5}^{+45.4}$ & $610.5_{-44.3}^{+28.2}$ \\ 
& & & & \\[-0.20cm] \tableline & & & &\\[-0.20cm]
1:1 E + E & e & $28.04_{-17.8}^{+49.7}$ & $0_{-0}^{+0}$ & $0_{-0}^{+0}$ \\
1:4 E + E & e & $1219_{-43.1}^{+28}$ & $441.6_{-2.45}^{+36.1}$ & $327_{-49.3}^{+43.3}$ \\
1:10 E + E & e & $916.8_{-4.87}^{+52.7}$ & $6.087_{-6.09}^{+5.7}$ & $0_{-0}^{+0}$ \\ 
& & & & \\[-0.20cm] \tableline & & & &\\[-0.20cm]
1:10 E + disk & e & $3431_{-44.9}^{+79}$ & $1344_{-32.8}^{+61.2}$ & $650.3_{-48.2}^{+52.7}$ \\
& & & & \\[-0.20cm] \tableline
\end{tabular}
\tablecomments{
The first column denotes the mass ratio and the type of the merger progenitors. 
The single letter in the second column specifies whether the binary system is initially bound and
the orbit is eccentric (e) or the orbital energy is zero and the orbit is initially parabolic (p).
The error limits correspond to lower and upper quartiles derived by adding random values with mean zero and a 
spread determined from the lower and upper quartiles shown in Figure \ref{fig:tidalParam} to the original 
tidal parameter curve and measuring the tidal-lifetime for each randomization. 
The values of $\theta$ corresponding to weak and strong features are chosen in a way such that weak 
features are barely noticeable under visual inspection of the mock image while strong features are 
outstanding emergences of tidal distortions.}
\end{center}
\end{table*}

The morphological properties of merger progenitors and remnants are analyzed from the projected images. Using the 
\texttt{ellipse} task of IRAF the ellipticity $\epsilon$ and the isophotal shape $100a_4/a$ are 
determined as a function of semi-major axis length $a$. The effective radius is estimated as a product between 
semi-major axis length $a_h$ containing half the light and $(1-\epsilon_h)^{1/2}$, where $\epsilon_h$ is the 
ellipticity measured at $a_h$. The central velocity dispersion is measured within a circular aperture of 1kpc radius 
mimicking observations~\citep{1992ApJ...399..462B,1987ApJS...64..581D}. In Table~\ref{tab:kinStrucProperties} we list the median structural and kinematic properties averaged in the range of 0.8 - 1.2 times the half-light semi-major axis
as obtained from at least 50, but typically more than 200 random, projections.

\begin{table*}[htbp]
\begin{center}
\caption{The Kinematic and Structural Properties of the Massive Elliptical and the Merger Remnants\label{tab:kinStrucProperties}}
\begin{tabular}{ccccccc}
\tableline
\tableline
& & & & & &\\[-0.20cm]
& & $M_e$ & $r_e$ & $\sigma_c$ & & \\ 
Description & Orbit & ($10^{10}M_\odot$) & (kpc) & (km s$^{-1}$) & $\epsilon$ & $100a_4/a$ \\ 
& & & & & &\\[-0.20cm]\tableline & & & & & &\\[-0.12cm]
Massive elliptical (E) & --- & 13.84 & $9.24_{-0.65}^{+0.23}$ & $248_{-9.8}^{+14}$ & $0.383_{-0.098}^{+0.039}$ & $-2.58_{-0.47}^{+1.5}$  \\ 
& & & & & &\\[-0.20cm]\tableline & & & & & &\\[-0.20cm]
1:1 E + E & p & 27.69 & $16.8_{-1.6}^{+1.5}$ & $296_{-21}^{+27}$ & $0.47_{-0.16}^{+0.079}$ & $-2.41_{-0.46}^{+1.5}$  \\
1:4 E + E & p & 17.3 & $13.9_{-1.1}^{+0.57}$ & $277_{-15}^{+31}$ & $0.341_{-0.077}^{+0.061}$ & $0.585_{-0.43}^{+0.33}$  \\
1:10 E + E & p & 15.23 & $11.8_{-0.37}^{+0.49}$ & $283_{-6.7}^{+6.1}$ & $0.261_{-0.057}^{+0.043}$ & $-0.97_{-0.46}^{+0.34}$  \\ 
& & & & & &\\[-0.20cm]\tableline & & & & & &\\[-0.20cm]
1:4 E + disk/bulge, rperi 9kpc & p & 17.7 & $13.2_{-0.79}^{+0.61}$ & $266_{-12}^{+19}$ & $0.325_{-0.086}^{+0.053}$ & $0.133_{-0.59}^{+0.74}$  \\
1:4 E + disk/bulge & p & 17.7 & $12.8_{-0.82}^{+0.79}$ & $266_{-13}^{+19}$ & $0.369_{-0.083}^{+0.061}$ & $0.454_{-0.29}^{+0.34}$  \\
1:4 E + disk/bulge, rperi 36 kpc & p & 17.7 & $12.9_{-0.95}^{+0.67}$ & $267_{-12}^{+20}$ & $0.391_{-0.088}^{+0.057}$ & $0.64_{-0.31}^{+0.43}$  \\
1:4 E + disk & p & 17.3 & $14.5_{-0.88}^{+0.89}$ & $264_{-5.8}^{+11}$ & $0.383_{-0.086}^{+0.06}$ & $0.325_{-0.29}^{+0.38}$ \\
1:10 E + disk/bulge & p & 15.22 & $11.1_{-0.53}^{+0.57}$ & $277_{-18}^{+22}$ & $0.305_{-0.091}^{+0.044}$ & $0.789_{-0.48}^{+0.43}$  \\
1:10 E + disk & p & 15.21 & $11.6_{-0.55}^{+0.59}$ & $276_{-6.3}^{+11}$ & $0.245_{-0.062}^{+0.056}$ & $-0.854_{-0.31}^{+0.35}$  \\ 
& & & & & &\\[-0.20cm]\tableline & & & & & &\\[-0.20cm]
1:1 E + E & e & 27.69 & $14.8_{-1.5}^{+1.8}$ & $289_{-12}^{+24}$ & $0.341_{-0.14}^{+0.13}$ & $0.0515_{-0.47}^{+0.56}$  \\
1:4 E + E & e & 17.3 & $11.6_{-0.61}^{+0.91}$ & $287_{-21}^{+37}$ & $0.288_{-0.064}^{+0.082}$ & $0.759_{-0.45}^{+0.4}$  \\
1:10 E + E & e & 15.23 & $10.8_{-0.66}^{+0.73}$ & $277_{-14}^{+16}$ & $0.257_{-0.06}^{+0.064}$ & $-0.757_{-0.51}^{+0.52}$  \\ 
& & & & & &\\[-0.20cm]\tableline & & & & & &\\[-0.20cm]
1:10 E + disk & e & 15.21 & $10.6_{-0.59}^{+0.75}$ & $270_{-9.3}^{+12}$ & $0.24_{-0.082}^{+0.06}$ & $-0.277_{-0.27}^{+0.29}$ \\
& & & & & &\\[-0.20cm]\tableline
\end{tabular}
\tablecomments{The first three columns denote the merger progenitor or remnant, respectively, the orbital setup (parabolic or eccentric), and half of the total system's mass. The values of
the following columns express the median values over random projections. In particular, the effective radius, 
the central velocity dispersion within 1 kpc, the ellipticity at effective radius, and the isophotal shape are given.
Error limits refer to lower and upper quartiles. The properties of the mass-rescaled versions of the 
elliptical progenitor (E) can be obtained from the scaling relations for the coordinates and velocities.}
\end{center}
\end{table*}

\begin{figure*}[htbp]
\begin{center}
\includegraphics[width=140mm,bb=35 134 536 654]{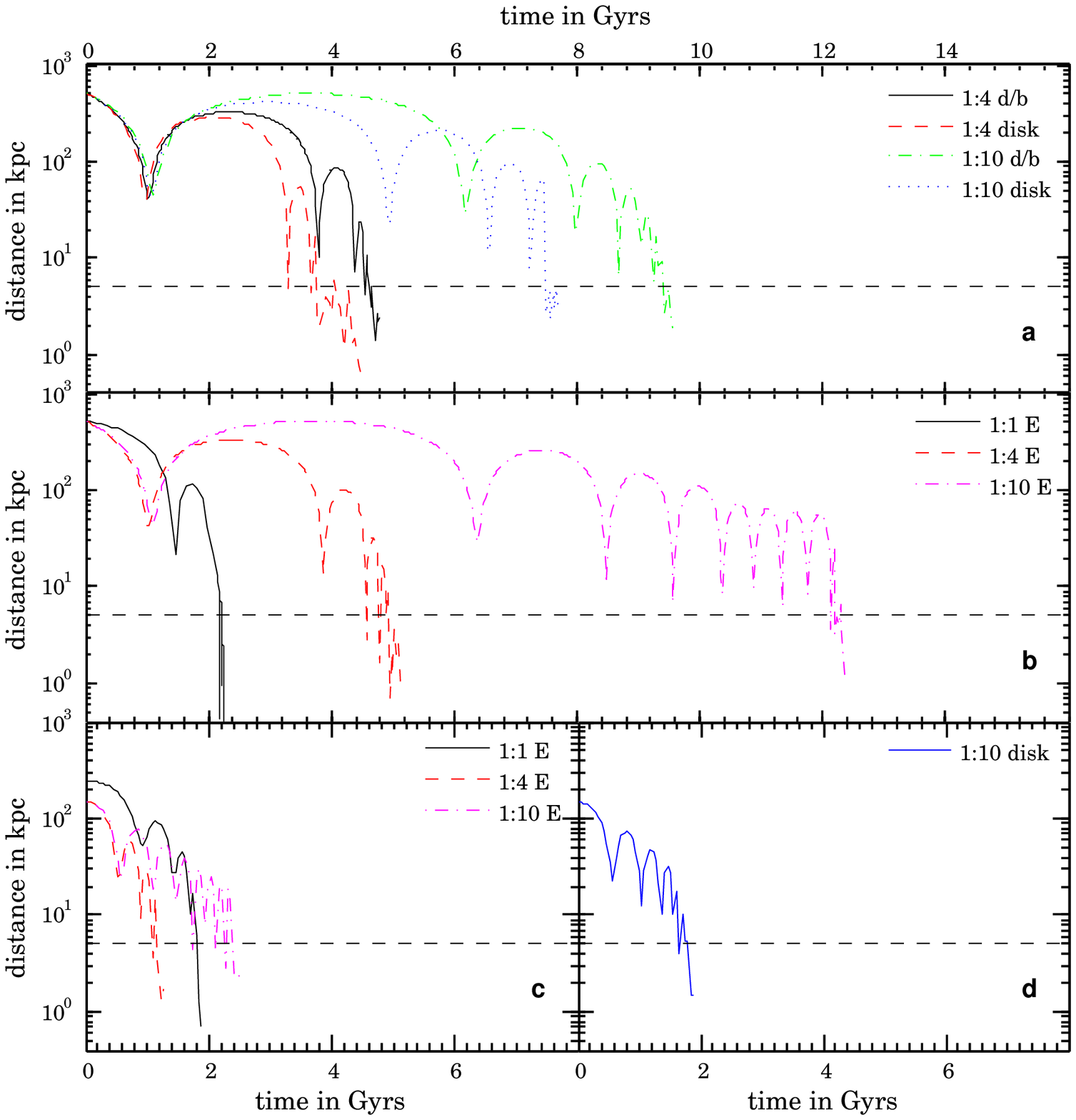}
\caption{Distance of the star particles with highest local density from each progenitor as function of time.
The first time that the latter distance drops below 5kpc and (except for noise) remains below this
threshold defines the merging time. 
Note that mergers of a given mass ratio between their stellar components but with different 
satellite type may have different total mass ratios (the total mass including dark matter). 
($a$) Mergers on parabolic orbits between a massive elliptical and pure disk satellites (disk) or satellites with 
disk and bulge (d/b).
($b$) Mergers between elliptical galaxies (E) on parabolic orbits.
($c$) Mergers between elliptical galaxies on eccentric orbits.
($d$) Merger between a massive elliptical and a 10 times less massive disk satellite on an eccentric orbit.}
\end{center}
\label{fig:orbits}
\end{figure*}

We define the merger time based on the particle data.
In particular, we keep track of the distance between the star particles with the highest density of 
each merger constituent and define the time of merging as the time when this distance drops below 5 kpc
(Fig.~\ref{fig:orbits}).

The overall merging time is shorter for more massive companion galaxies. It is determined
by a combination of the dynamical friction time with the timescale of mass loss via tidal stripping~\citep{2003MNRAS.341..434T}. 
The dynamical friction timescale is longer for a lighter companion and tidal stripping slows it 
down further~\citep{1999ApJ...525..720C,2003MNRAS.341..434T}. More than half of the mass of the dark halo is 
typically stripped along the first orbit, while the stellar component loses mass 
more gradually since it sits deeper in the potential well of the halo.
When measured from first pericenter the merger takes on parabolic orbits $<1$ Gyr for equal mass galaxies, and between $7$ and up to $12$ Gyr 
for a 1:10 mass ratio. The merging time is generally faster on eccentric orbits because they start with a 
smaller orbital time in our initialization, which yields a shorter dynamical friction time. 
Finally, pure disk companions are
disrupted much faster than disks with bulges or elliptical companions due to their lower central density,
which results in a smaller tidal radius and a more impulsive response to the tidal shocks at pericenters~\citep{1999ApJ...514..109G}.

An observer would define a merger as complete when the individual merger components could not be distinguished anymore. Using
our mock observational setup we find that equal mass mergers happen on such a fast timescale that the particle-based
merging time and the observational merging time coincide. A similar statement holds for the simulations 
involving only elliptical galaxies since in this case the core remains extremely compact and reaches the
inner kiloparsec of the host galaxy without being destroyed. For our 1:10 merger involving a pure disk on the other hand there is a significant
difference of $\sim{}200$ Myr; i.e. the progenitors have not yet merged according to the particle definition, but the 
satellite is so strongly disrupted that it is difficult to recognize it as a single individual object. 

\section{Results}
\subsection{Strength and lifetime of tidal features}
\label{sect:tidalStrength}

\begin{figure*}[htbp]
\begin{center}
\includegraphics[width=120mm, bb=82 184 490 592]{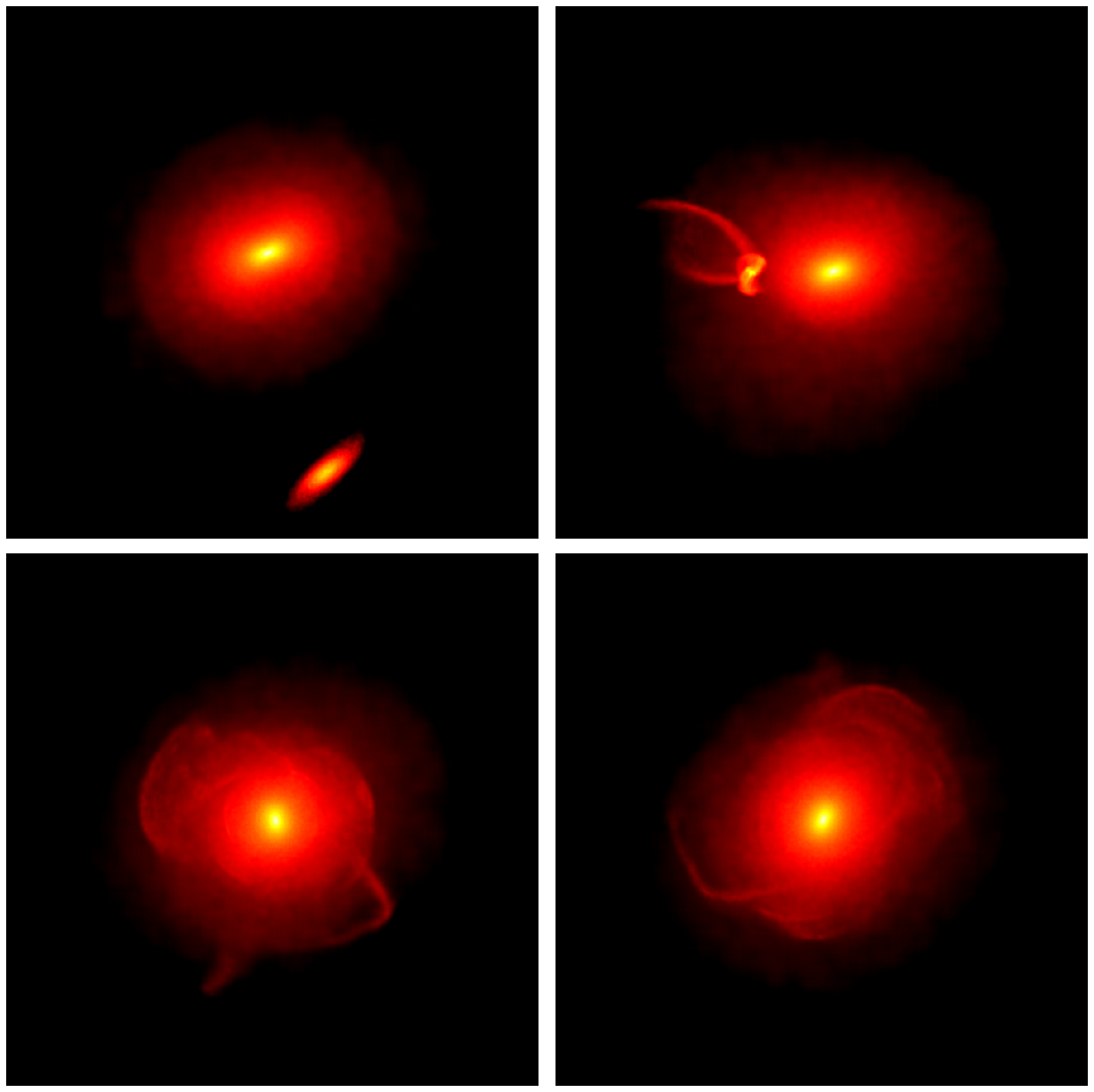}
\caption{Evolution of merging galaxies. Surface brightness maps showing the stellar component at different
stages of the simulation of a merger between a massive elliptical galaxy and a bulgeless disk 
with 1/10 of the mass of the accreting elliptical; the systems are on an eccentric orbit. Top: Merging systems 200 Myr after the beginning of the simulation (left) and 800 Myr later on their way to the second pericentric passage (right). At this point the disk is already heavily damaged and tidal streams appear. Bottom: Remnant 500 Myr (left) and 1 Gyr (right) after the merger. The visible tidal streams stem from the disk and begin to develop at the first pericentric passage. They survive up to a few Gyr after the merger. All images have a physical size of 300$\times{}$300 kpc. The surface brightness ranges from 28.5 mag arcsec${^{-2}}$ (black) to 18 mag arcsec${^{-2}}$ (white). Noise is not added to these images.}
\end{center}
\label{fig:evolution}
\end{figure*}

Fig.~\ref{fig:evolution} shows an example of the dynamical evolution of one of the elliptical-disk merger simulations. The mass ratio between the satellite disk and the main elliptical is 0.1. Stars belonging to the disk of the small satellite are stretched in tidal streams  at the first pericentric passage. These tidal features, which originate from the companion's disk component, survive for a few Gyr after the merger, in contrast with the short-lived debris that is produced in elliptical-elliptical mergers. We quantify the strength of the tidal features by means of the tidal parameter $\theta$; this measures the light excess  in the tidal tails relative to a smooth elliptical-isophote model of the galaxy, see section \ref{sect:analysis}.

Mergers between elliptical galaxies, bona fide dry mergers, produce tidal features that are short lived; in particular equal-mass mergers do not show strong features for more than 100 Myr (Table~\ref{tab:tidalTimes}).
Mergers with disk-dominated satellites, instead, produce not only stronger but also longer lived features.  Strong, easily identifiable features 
($\theta>0.15$) last for more than 1 Gyr, while weak, barely visible features ($\theta\approx{}0.1$) persist for up to 3 Gyr.
We explain this difference with the smaller velocity dispersion of the disk material, which makes up for most of the stellar tidal debris even in the galaxies with a bulge component, which implies that phase mixing with the background mass distribution occurs on a longer timescale and that tidal features are thus longer-lived compared to dry mergers. 

\begin{figure*}[htbp]
\begin{center}
\includegraphics[width=160mm]{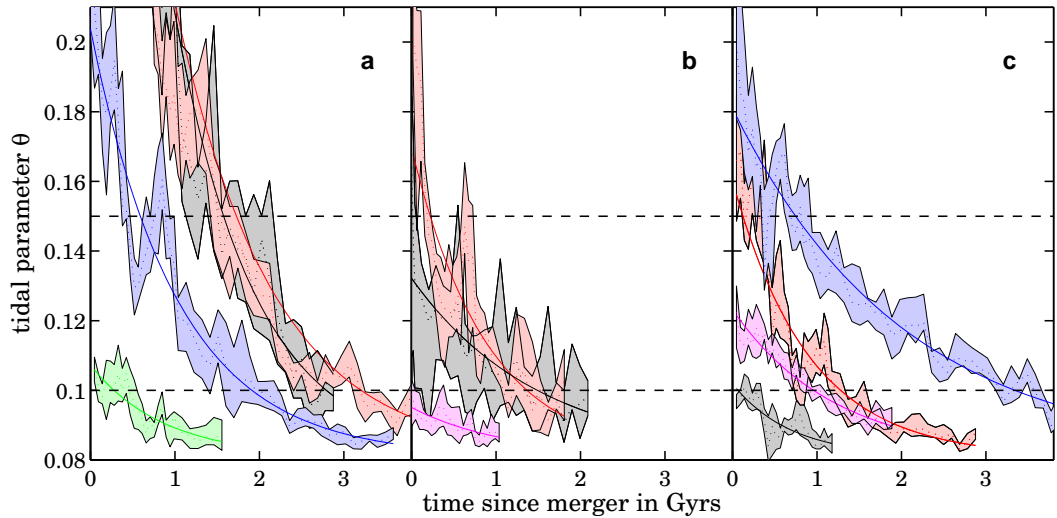}
\caption{Tidal parameter $\theta$ as function of time elapsed after the merger.
For each merger set-up and time step, the tidal parameter is calculated over 20 random projections. The colored 
dashed lines show its median values, and the shaded area defines the lower and upper quartile. Colored dotted lines indicate exponential fits to the declining tidal parameter.
($a$) Parabolic encounters involving disk galaxies. Shown are a 1:4 merger with a pure disk satellite (red), a 1:4 merger with  disk+bulge satellite (black), a 1:10 merger with a pure disk satellite (blue), and a 1:10 merger with a disk+bulge satellite
(green). The strongest features ($\theta>0.15$) are visible for the most massive disks and, in this example, last for about 1.5 Gyr. 
($b$) Dry mergers between elliptical galaxies on parabolic orbits with stellar mass ratios 1:1 (black), 1:4 (red) and 1:10 (magenta). Strong features are visible for only a very short time. 
($c$) Encounters on eccentric orbits. Black, red, and magenta curves denote elliptical-elliptical mergers as in ($b$). In addition, a 1:10 merger with a disk galaxy is shown (blue).}
\end{center}
\label{fig:tidalParam}
\end{figure*}

We find that the time dependence of the tidal parameter can be well fit with a declining exponential
of the form $\theta(t)=Ae^{-t/\tau}+B$ (Fig.~\ref{fig:tidalParam}). The additive constant $B$ is kept fixed at $B=0.085$, which is the value that
we measure for our elliptical galaxy model in isolation and only $A$ and $\tau$ are fitted. The product 
of $A$ and $\tau$, which represents the area under the curve of the tidal 
parameter once the offset $B$ is removed and is hence a measure of the strength of the tidal features, 
discriminates well between dry mergers and mergers involving disks. Each of our dry merger simulations shows $A\tau \lesssim{} 0.07$ Gyr while 
all our disk simulations have $A\tau \gtrsim{} 0.1$ Gyr, with the notable exception of the 1:10 merger involving a disk+bulge system.

The strength of tidal features originating from a merger with a pure disk satellite and from a merger 
with a disk plus bulge satellite are found to be of comparable strength for a mass ratio of 1:4 but 
very different for a mass ratio of 1:10. In particular, the merger with a disk plus bulge system 
does not show strong tidal features at merging time as opposed to the strong and long-lived features in the 1:10 merger
with a pure disk satellite. We explain this difference as follows. In our simulation the bulge-less 
satellite is disrupted before properly sinking to the center of the giant elliptical. Hence, the merging time, which is 
determined by the minimum of disruption time and tidal friction time, is set by the smaller of the two, which in this case is the tidal disruption time. On the other hand galaxies containing a bulge are more concentrated and hence less susceptible to 
both tidal stripping and tidal heating. This in effect increases the tidal disruption time and consequently the merging time, which is now essentially determined by the dynamical friction time~\citep{2003MNRAS.341..434T}.
The concentrated, 
cold tidal features in our simulations originate from the stellar disk and not from the kinematically hot bulge stars. The features
are mainly created at pericentric passages and at the final coalescence of the disk component with the elliptical galaxy (as tidal forces scale 
with the cube of the inverse distance). Hence, the presence of a bulge introduces a significant time span between the creation 
of the tidal features and the final merger of the dense core of the satellite with the center of the giant elliptical. The produced
tidal features may thus become severely dimmed or even invisible at merging time. In the case of 
the 1:4 mass ratio this time span is small for both the satellites with bulge and without 
because for companions of such a high mass the dynamical friction timescale is
always dominant in setting the merging time (the dynamical friction time is only mildly 
sensitive to the internal structure of the secondary galaxy).

We also study the dependence of the tidal debris on the orbital parameters of the encounter for
a given structure and mass ratio of the two galaxies.
Rerunning the 1:4 merger involving a disk plus bulge satellite on a parabolic orbit with twice and with half the Keplerian pericenter of 18 kpc 
does not change the tidal strength and life-times significantly. 
For eccentric bound orbits, which is the more natural assumption if the merger occurs in a virialized group rather than in the field, tidal 
debris lasts slightly longer than for the corresponding mergers on parabolic orbits. The reason is that disruption occurs deeper
in the central part of the potential of the primary for such eccentric orbits. Closer to the
center of our elliptical galaxy model we find the inertia tensor to be less triaxial and the effect of orbital precession, which 
wraps around and phase mixes tidal streams in triaxial systems~\citep{2002MNRAS.336..119M}, is reduced.

\subsection{Morphology of the tidal debris}
\label{sect:tidalMorph}

\begin{figure*}[htbp]
\begin{center}
\includegraphics[width=120mm, bb=82 184 490 592]{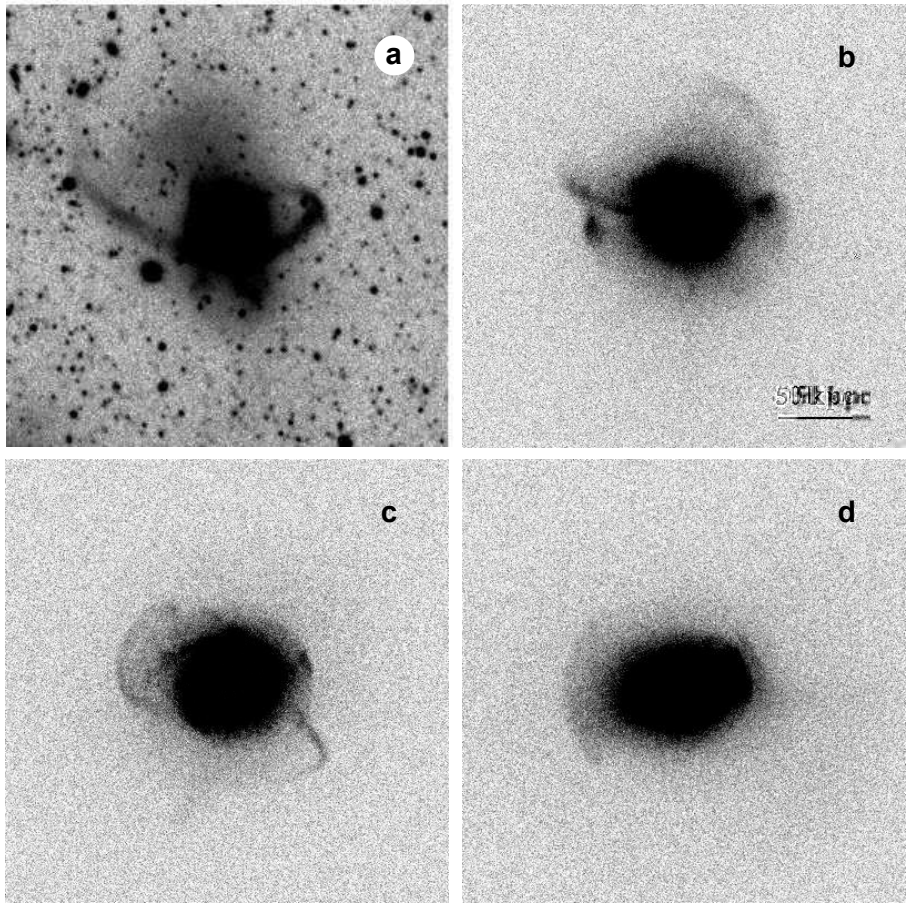}
\caption{Morphological signatures of mergers. 
($a$) Observed image of an elliptical galaxy that shows red tidal tails at large galactocentric distances~\citep{2005AJ....130.2647V}. 
($b$) Mock image of a simulated 1:4 merger between an 
elliptical galaxy and a disk+bulge system; the snapshot is taken 600 Myrs after the merger. 
($c$) Mock image of a simulated 1:10 merger between an elliptical galaxy and a disk; the snapshot is taken 500 Myrs after the merger. 
($d$) Mock image of a simulated 1:4 merger between \emph{elliptical} galaxies, 380 Myrs after the merger. 
There is a striking difference between the tidal features originating in mergers between ellipticals and  those involving a disk-dominated companion, independent of the mass ratio and orbit characteristics. While all mergers can lead to shells and diffuse features, only the mergers involving disks show strong tidal arms and loops, similar to the observed features around bright ellipticals (see [$a$]).
Panel $a$ has a size of 2\farcm{}5$\times{}$2\farcm{}5, corresponding to a physical extension of about 
280$\times{}$280 kpc for an object at redshift $z\sim{}0.1$. The mock images ($b$-$d$) have a physical dimension of 300$\times{}$300 kpc and have been generated with resolution and noise properties so as to match the observational data used for the comparison~\citep{2005AJ....130.2647V}.
The bright internal regions of the images are saturated, in order to increase the contrast on the tidal features.}
\label{fig:morphology}
\end{center}
\end{figure*}

A striking difference between elliptical-elliptical and elliptical-disk mergers is manifested in the morphology of the resulting tidal tails (Fig.~\ref{fig:morphology}). Mergers between two kinematically hot early-type systems produce diffuse plumes of stars, which vanish rapidly \citep{2005AJ....130.2647V,2006ApJ...640..241B}.
In some of our simulated mergers between early-type galaxies of \emph{unequal} mass we observe in addition straight, elongated tails, which also vanish rapidly. On the other hand, we find that mergers that involve a dynamically cold disk are able produce strong and long-lived tidal features, with loops and arms resembling those observed around bright ellipticals~\citep{2005AJ....130.2647V}.

In order to estimate the fraction of systems with broad tidal features versus features with loops and arms
we visually classify the morphology of the tidal debris in the sample of 86 nearby bulge-dominated red 
field galaxies~\citep{2005AJ....130.2647V}. Twenty-five of the selected galaxies are classified as possessing no observable tidal features,
27 objects are classified as showing weak features, 14 objects show strong features, and 20 objects are currently 
undergoing a merger. For six galaxy pairs the merger companions are included in the sample of the 20 merging galaxies, leaving 17 unique red merger pairs.
We find that of 57 unique sample galaxies for which signs of interactions are detected (out of the total sample of 83=86-3 
bulge-dominated, red field galaxies), only nine systems show the ``broad fans'' that are expected from a very recent elliptical-elliptical merger. In contrast, 19 systems clearly show tidal arms and loops.
This statistics  should  of course be taken with care, in the light of the 29 cases in which the tidal features  are too faint for a decisive classification of their morphology.
Hence, our qualitative visual inspection confirms that the majority of the observed tidal features in elliptical galaxies are indeed consistent 
with mergers involving disk satellites rather than bona fide dry mergers.

\begin{figure}[htbp]
\begin{center}
\includegraphics[width=80mm, bb=130 251 444 520]{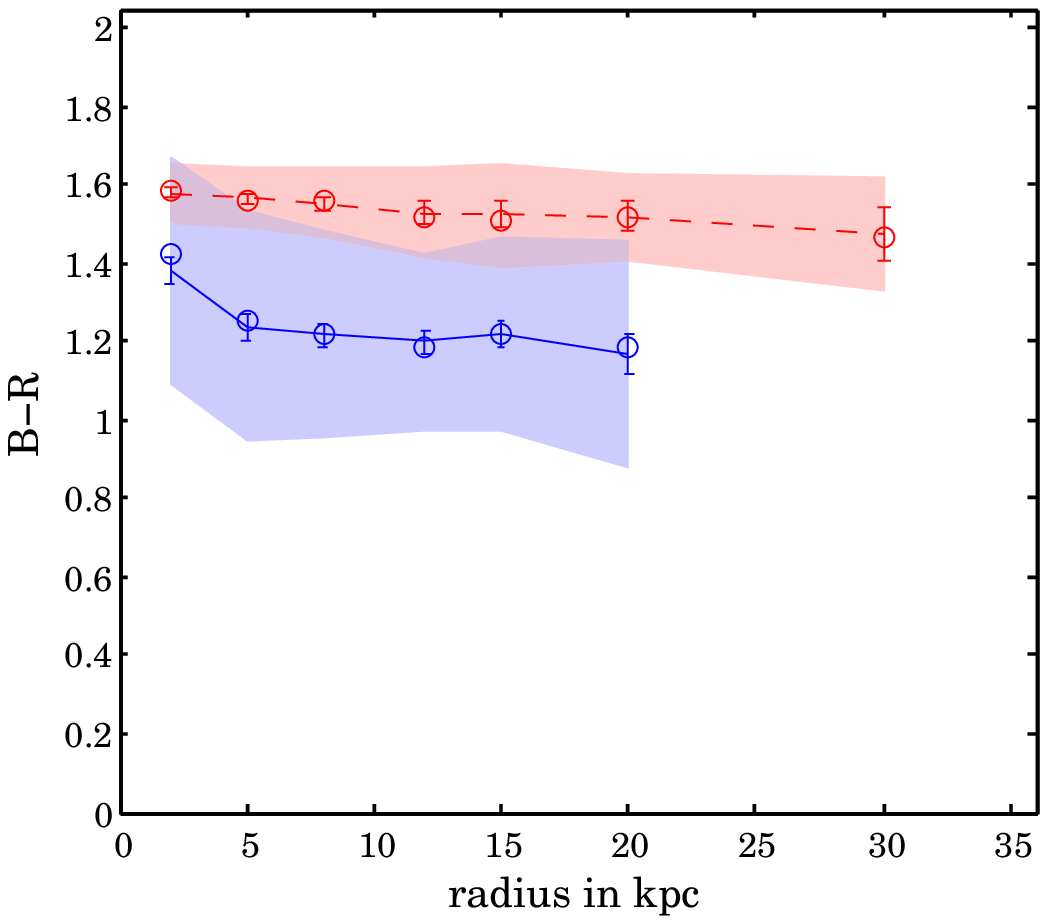}
\caption{Dependence of the $(B-R)$ color on radius measured in physical kiloparsecs.
Lines denotes the average $(B-R)$ color in the radial bin of the respective sample of spiral galaxies (\citealt{1994AaAS..106..451D}, blue solid line) and giant elliptical galaxies (\citealt{1990AJ....100.1091P}, red dashed line). Error bars denote the 1 $\sigma$ error of the average $(B-R)$ color. Circles indicate the median $(B-R)$ color in the respective radial bin. The shaded area shows the 1 $\sigma$ scatter for individual color measurements.}
\label{fig:deJongPeletierkpc}
\end{center}
\end{figure}

\begin{figure*}[hbtp]
\begin{center}
\begin{tabular}{cc}
\includegraphics[width=80mm, bb=142 251 450 520]{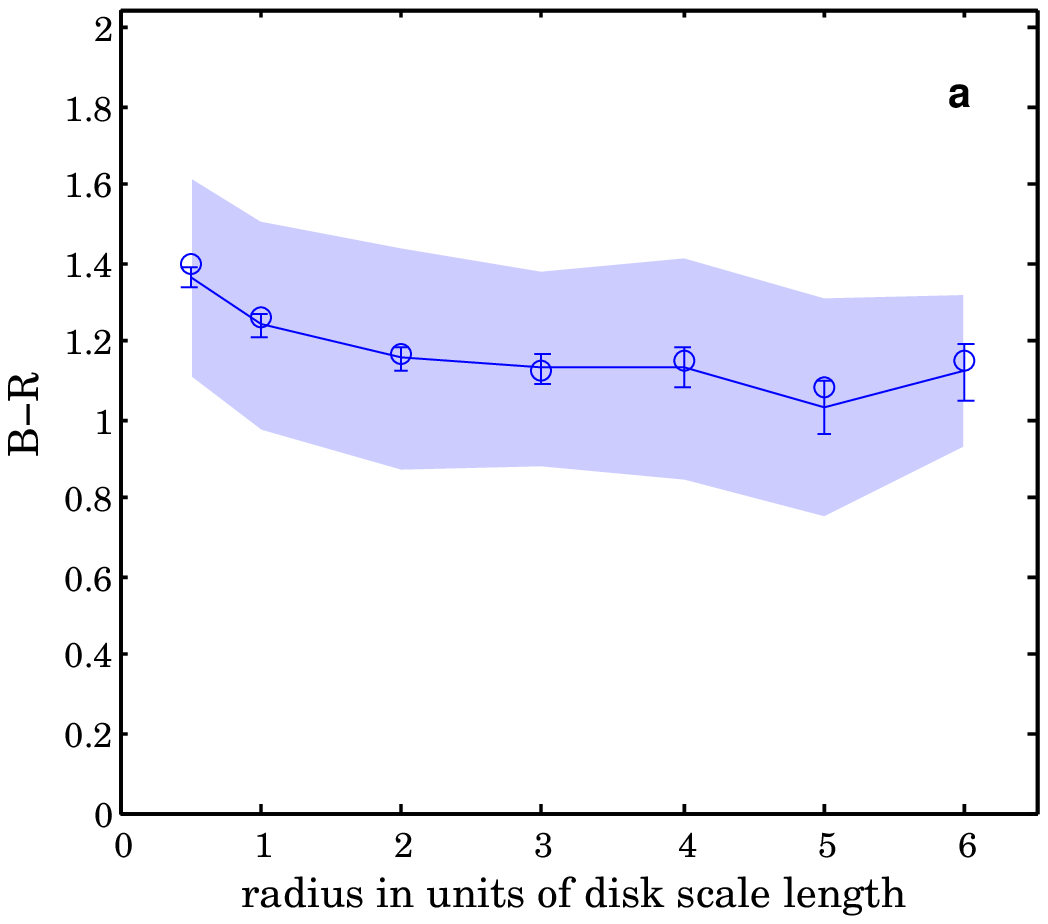} & \includegraphics[width=80mm, bb=142 251 450 520]{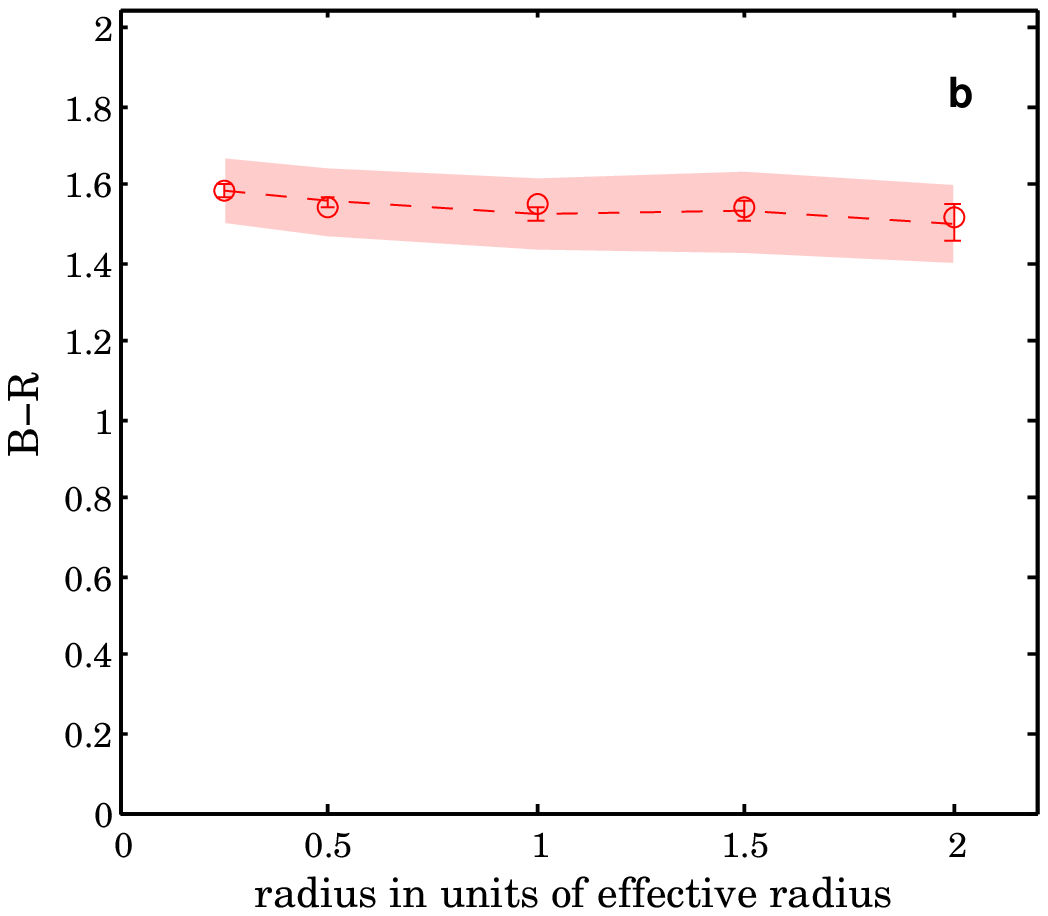} \\
\end{tabular}
\caption{Dependence of the $(B-R)$ color on radius measured in units of galaxy's scale lengths.
($a$) The $(B-R)$ color of spiral galaxies~\citep{1994AaAS..106..451D} as function of radial disc scale length. Disc scale lengths are derived by a linear fit to the surface brightness curve between 10 and 50 arcsec. 
($b$) The $(B-R)$ color of giant elliptical galaxies as function of half-light  radius (data from~\citealt{1990AJ....100.1091P}). The major axes provided in the original reference have been transformed  to half-light radii using the ellipticity at 10 arcsec. Symbols and shaded areas are as in Fig.~\ref{fig:deJongPeletierkpc}.}
\label{fig:deJongPeletierScale}
\end{center}
\end{figure*}
\subsection{The color of the tidal features and the merger remnant}
\label{sect:tidalColor}

In the following discussion we assume Johnson $B_\textrm{J}$ and Cousins $R_\textrm{C}$ filter bands. Magnitudes and colors are normalized to Vega.
Our simulations show that the tidal arms and loops stem mainly from the regions of the companion's disk at about  $2 - 5$ radial scale lengths from the center (i.e., at about  the solar neighborhood in the Milky Way).
The observed $(B-R)$ color of M31 at such distances, for example, is about 1.4 mag~\citep{1988A&A...198...61W,2007arXiv0707.4375T}.\footnote{The stellar mass ratio between M31 and our elliptical galaxy model is about 1:4. Less massive disks will be on average less luminous, more gas-rich, and bluer.} This is only $\sim 0.1$ mag bluer than the typical color of giant ellipticals at 30-50 kpc from the center, where the tidal tails are observed. Indeed,  the cores of nearby giant ellipticals have typically $(B-R)\sim1.6$ mag, (e.g.~\citealt{1999A&AS..137..245M}), and, given the typical color gradients that are observed for giant ellipticals~\citep{1990AJ....100.1091P}, the color at 30-50 kpc is typically  $(B-R)\sim1.5$ mag (Fig.~\ref{fig:deJongPeletierkpc}).

To stress this point further,  we compile in 
Fig.~\ref{fig:deJongPeletierScale}a the $(B-R)$ colors as a function of 
scale length from the sample of~\citet{1994AaAS..106..451D}; the sample contains 86 face-on nearby disk-dominated galaxies.  We note that $(a)$ the typical $(B-R)$ color of disk galaxies between 2 and 5 radial scale lengths is $\sim{}1.2$ mag; and $(b)$ the 1 $\sigma$ fluctuations for individual galaxies are of the order of 0.25 mag.
For a direct comparison, we show in Fig.~\ref{fig:deJongPeletierScale}b
the $(B-R)$ color of the sample of 39 giant elliptical galaxies of~\citet{1990AJ....100.1091P}.  At about two half-light radii, i.e., at the distance at which tidal loops and arms are observed, the typical $(B-R)$ color of ellipticals can be seen to be $\sim1.5\pm0.1$ mag. Hence, the $(B-R)$ color difference between the disk stellar population responsible for the tidal arm and the background population of an  elliptical galaxy is of the order of 0.3 mag,  and its typical scatter is of the same order.
This difference in colors of a few tenths of a magnitude is already quite
small but has to be viewed as a conservative number. In fact, the disk galaxies in the sample of~\citet{1994AaAS..106..451D} are normal star-forming disk galaxies, while in our case the disk satellite will have the star formation quenched in the outer region as soon as it completes the first close approach with the primary elliptical as a result of gas stripping. As seen in
hydrodynamical simulations tides will
remove the gas at least down to the same radius at which stars are removed (which is down to 2.5 disk scale lengths
in our simulations) and ram pressure in the gaseous
halo that the giant elliptical will presumably possess will enhance stripping even further~\citep{2006MNRAS.369.1021M}.
Tidal tails originating from 2-5 disk scale lengths would thus be indeed {\it red} -- not blue -- features.

\begin{figure}[htbp]
\begin{center}
\includegraphics[width=80mm, bb=136 251 446 525]{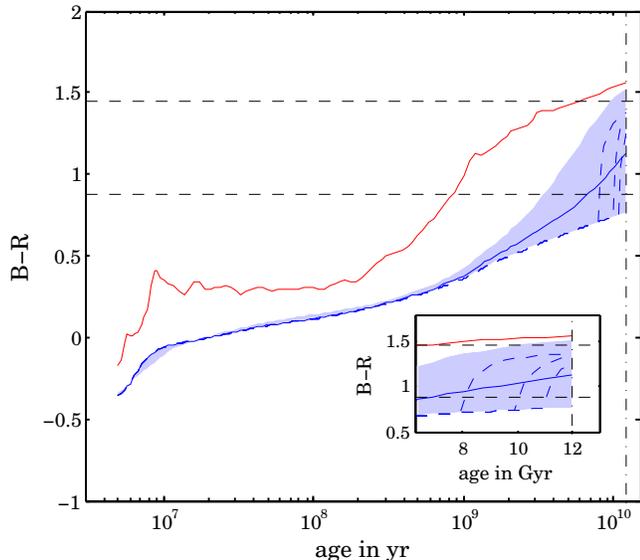}
\caption{Evolution of $(B-R)$ color in synthetic stellar population models~\citep{2003MNRAS.344.1000B}.
The $(B-R)$ color as function of age for a simple stellar population of solar metallicity (red solid line). The blue solid line shows the $(B-R)$ color of a composite stellar population with exponential declining star formation rate, a 4 Gyr $e$-folding time, and half-solar metallicity. The shaded area brackets the $(B-R)$ colors once we vary the $e$-folding time between 1.5 Gyr and infinity, i.e. a constant star formation rate, and the metallicity between half-solar and solar. The colors are assumed to be measured 12 Gyr after the initiation of the star formation (vertical dot-dashed line). The horizontal dashed lines indicate the lower and upper quartiles in $(B-R)$ color at 3 disc scale lengths from the center of the disk (from Fig.~\ref{fig:deJongPeletierScale}a). Dashed blue lines indicate the $(B-R)$ color of a stellar population with initially constant star formation that is then quenched 4, 2 or 1 Gyr before the observation. The inset shows the $(B-R)$ color evolution restricted to the last 6 Gyr previous to observation. Our models assume a Chabrier initial mass function \citep{2003PASP..115..763C}.}
\label{fig:DBR}
\end{center}
\end{figure}

This conclusion is strengthened by an analysis conducted with stellar population synthesis models~\citep{2003MNRAS.344.1000B}.
In Fig.~\ref{fig:DBR} 
we plot the $B-R$ color as a function of epoch for simple stellar populations (red solid line). For our elliptical of mass $\sim{}3\times{}10^{11}$ $M_\odot$ we assume overall solar metallicity~\citep{1998A&A...334...99K}.
We note that the $(B-R)$ color varies between -0.2 and 1.6 mag, i.e., it spans more than 1.8 mag, when moving from very young ($<10^7$ yr) to very old (12 Gyr) stellar populations. We model the $(B-R)$ color of a $\sim{}3\times{}10^{10}$ $M_\odot$ disk galaxy by assuming an exponentially declining star-formation rate with a 4 Gyr $e$-folding star formation timescale~(\citealt{1998ApJ...500...75L},\citealt{2000MNRAS.312..497B};blue solid line). We find that the $(B-R)$ color of our disk is typically only $\sim{}0.4$ mag smaller then the $(B-R)$ color of the bulk stars of the massive elliptical. We also consider a set of models with $e$-folding times ranging from 1.5 Gyr to infinity (i.e. constant star formation rate) and metallicities varying from half-solar to solar. These models predict a spread in $(B-R)$ color of about $0.4$ mag. The horizontal dashed lines indicate the lower and upper quartile of the $(B-R)$ color of face-on disks at about 3 disk scale lengths (Fig.~\ref{fig:deJongPeletierScale}a).
Models in which the star formation in the outer disk is suppressed as a result of gas stripping are also included (blue dashed lines). The considered timescale of quenching of star formation previous to observation of the tails ranges from 4 Gyr (upper blue dashed line) to 1 Gyr (lower blue dashed line). These models exhibit a $(B-R)$ color between 1.26 and 1.38 mag, i.e, only 0.1 - 0.25 mag less than the color of the bulk stars of the elliptical at the radius at which tidal tails are observed. Note that 1 Gyr is a conservative timescale, since material from the outer disk is stripped well before the merger is completed and the strong tails remain visible for up to 1.5 Gyr in disk-elliptical mergers. Our stellar synthesis models employ a Chabrier initial mass function~\citep{2003PASP..115..763C}.

We therefore conclude that in the qualitative analysis of~\citet{2005AJ....130.2647V} the colors of tidal debris originating from the disruption of a disk companion would be within at most a few tenths of magnitude from  the colors of the background stars of the host elliptical galaxy at those galactocentric distances and would thus  be identified  as ``red'' features. 

We note that in principle on-going star formation during and after the
completion of the merger could lead to tidal features with bluer colors. This
would weaken our conclusion that disk-elliptical minor mergers are
responsible for red tidal features. However, some observations suggest
that blue tails should exist only for a very short time after the merger
and should redden quickly \citep{1990AJ.....99..497S,1996AJ....111..655H}. Nevertheless, a further study
of this issue is necessary and will require the use of hydrodynamical
simulations with a realistic model of star formation.

\begin{figure}[htbp]
\begin{center}
\includegraphics[width=80mm,bb=100 209 496 559]{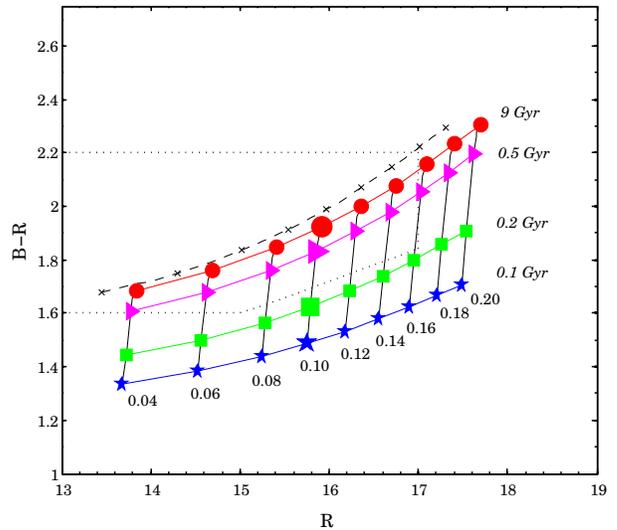}
\caption{Color-magnitude selection of luminous red galaxies. The selection window of~\citet{2005AJ....130.2647V}
is indicated by dotted lines. The dashed black curve corresponds to a 3$L$* galaxy~\citep{2001AJ....121.2358B}.
The solid lines show the expected colors and magnitudes as a function of time (0.1 - 9 Gyr) and 
for different redshifts (from 0.04 to 0.2) of an 
old stellar population of mass $2.7\times{}10^{11}$ $M_\odot$ after merging with a 10 times less massive disk. 
This disk shall contain 10\% of their mass in the form of gas, and we assume that about 70\% of the available gas in the 
disk is consumed in an exponentially declining starbursting phase. We use an $e$-folding time of 0.1 Gyr and an 
initial star formation rate of 20 $M_\odot$yr$^{-1}$ to parametrize the starburst~\citep{2005ApJ...623L..67K}. 
We find that 300 Myr after the beginning of the starburst the colors of the merger remnant are red enough to fall into the selection window of~\citet{2005AJ....130.2647V}.
The elliptical galaxy is modelled with a 9 Gyr old single stellar population with red horizontal branch
morphology~\citep{2005MNRAS.362..799M}. Both the old population and the starbursting population are assumed to have Salpeter initial mass functions 
and solar metallicity. All spectral energy distributions are derived from synthetic models~\citep{2005MNRAS.362..799M}. Magnitudes and 
colors are calculated with ZEBRA~\citep{2006MNRAS.372..565F} and normalized to Vega. Filter bands correspond to the Johnson-Cousins system.}
\label{fig:RedSelection}
\end{center}
\end{figure}

In Fig.~\ref{fig:RedSelection} we investigate the color of the massive elliptical after a 1:10 merger  
with a disk galaxy that contains 10\% of its mass in the form of gas. Within less than 300 Myr the B-R colors of the 
remnant are red enough to fall into the selection window of~\citet{2005AJ....130.2647V}.

\subsection{The fraction of tidal features originating from mergers with disks}
\label{sect:fraction}

For any given mass ratio, we calculate the average relative lifetimes $f_t$ of the \textit{weak} (Table~\ref{tab:tidalTimes}) tidal features generated respectively by elliptical-disk and elliptical-elliptical mergers (averaged over various simulation set-ups).  This represents a conservative lower limit on the relative lifetime, since stronger tidal features would have even larger
relative lifetimes.  We estimate that the average disk-elliptical-merger  lifetime  is a factor $\sim 2.3$ larger than the elliptical-elliptical-merger lifetime for a 1:4 mass ratio;  this factor becomes $>4$ for a 1:10 merger. 

\begin{figure*}[htbp]
\begin{center}
\begin{tabular}{cc}
\includegraphics[width=80mm, bb=124 242 466 533]{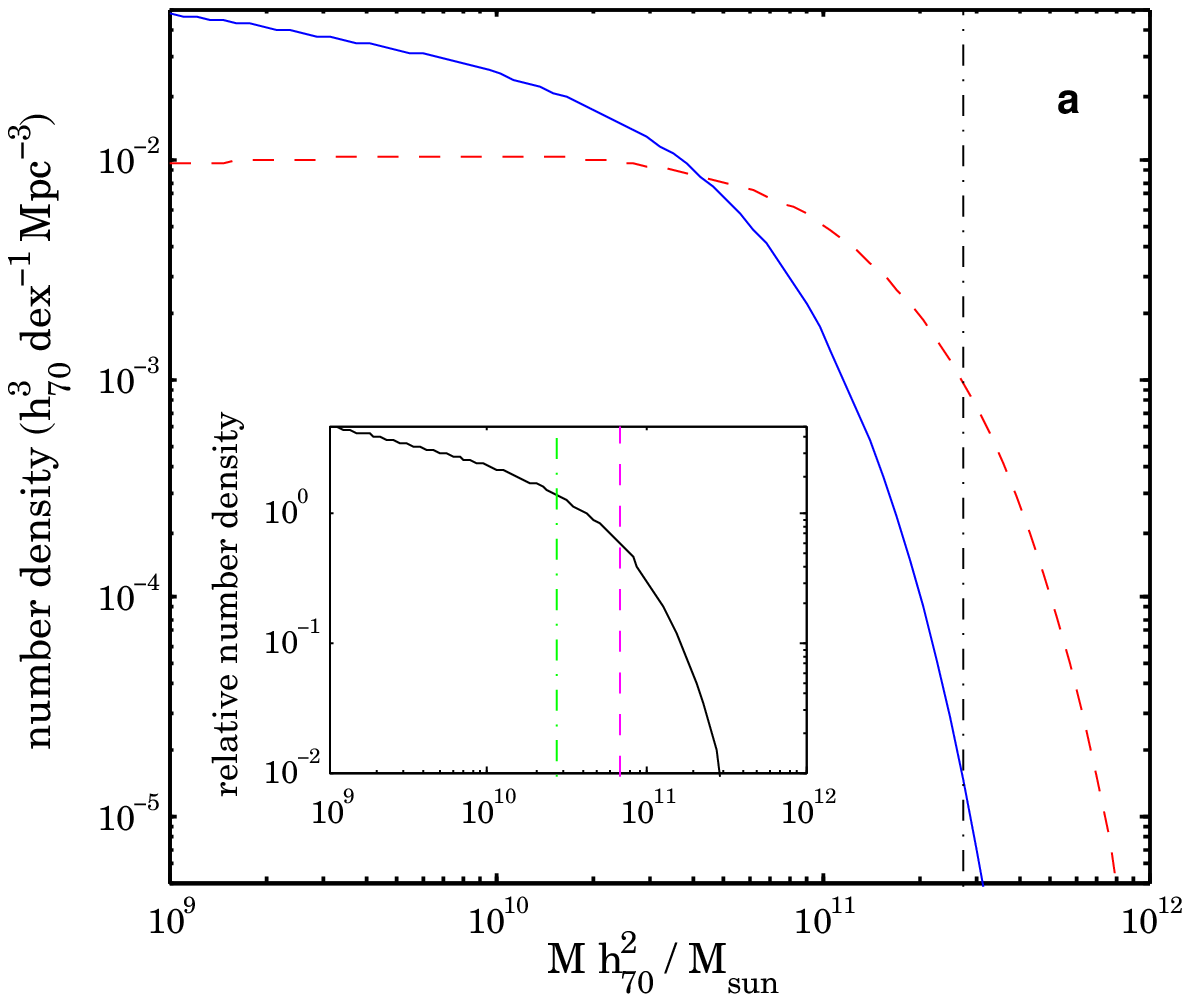} & \includegraphics[width=80mm, bb=124 242 466 533]{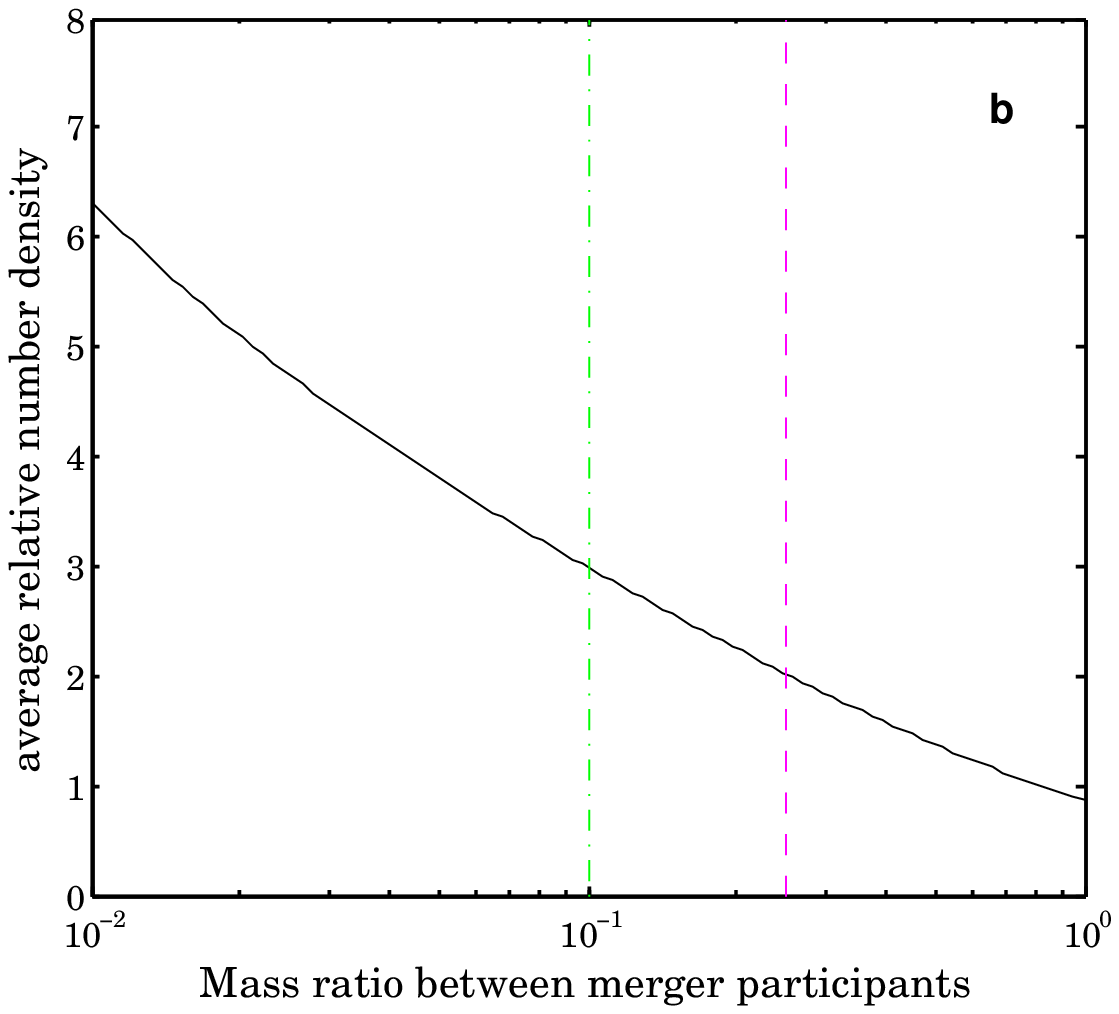} \\
\end{tabular}
\caption{(Relative) number densities of disk and early-type galaxies.
($a$) Stellar mass functions of early-type (red dashed line) and disk (blue solid line) galaxies in the local universe~\citep{2006MNRAS.368..414D}. The vertical, black dot-dashed line corresponds to galaxies of $2.7\times{}10^{11}$ $M_\odot$. The inset shows the ratio of the two stellar mass functions and indicates its value for a 4 times (magenta dashed line) and 10 times (green dot-dashed) less massive satellite galaxy. The ratio lies in the range of unity for the considered masses and increases with decreasing stellar mass.
($b$) Relative number density $f_n(R)$ of disk vs. early-type galaxies for a given mass ratio $R$ after integrating over all
masses of the elliptical host galaxy (black solid line). Mass ratios of 1:4 and 1:10 are indicated by the green dot-dashed and the magenta dashed lines. The respective values of the average relative number density are 2.0 and 3.0, respectively. Under the assumption that the merger probabilities are proportional to the number densities, a typical 1:4 or 1:10 merger 
with an elliptical galaxy will involve much more likely a disk than another early-type system.}
\label{fig:numberDens}
\end{center}
\end{figure*}

We use the $B$-band luminosity functions of~\citet{2006MNRAS.368..414D} to give a rough estimate of the relative mass fraction of early-type versus disk galaxies (see Fig.~\ref{fig:numberDens}a). 
Those authors classify by visual inspection galaxies into E/S0, Sa/b/c, or Sd/Irr/Pec. In the following analysis we identify the first class (E/S0) with early-type galaxies and the latter classes with disks. Our results do not depend on whether we include or exclude the Sd/Irr/Pec class because within our considered mass range such galaxies do not contribute significantly to the mass function. In order to obtain estimates for the mass functions near $\sim{}5\times{}10^{10}$ we divide the provided luminosity functions by a $B$-band mass-to-light ratio of 4.5 (E/S0), 2 (Sa/b/c), and 1 (Sd/Irr/Pec). These mass-to-light ratios are consistent with the considered range of $(B-R)$ colors derived from stellar synthesis modelling~\citep{2001ApJ...550..212B}.

Our elliptical galaxy host has a stellar mass of $2.7\times{}10^{11}$ $M_\odot$, which corresponds to $2.3M^*$  [with $M^*\sim{}1.2\times{}10^{11}$ $M_\odot$ provided by the Schechter function fit of~\citet{2006MNRAS.368..414D} to the luminosity function of early-type galaxies and assuming $(M/L)_B=5.5$]. For a mass ratio of 1:4 and 1:10, respectively,  we obtain a relative number 
density of disk versus early-type galaxies of 0.60 and 1.45 
(see inset of Fig.~\ref{fig:numberDens}a).
We have applied the same analysis to the mass functions derived by other authors who split galaxies according to different criteria.
From the mass functions of~\citet{2003ApJS..149..289B} we compute relative number densities of 0.64 and 1.32 for 1:4 and 1:10 mergers, respectively. The mass functions provided in~\citet{2004ApJ...600..681B} lead to relative number densities of 0.81 and 1.05.

We shall now assume that the relative merger frequency is proportional to the relative number density of disk versus early-type systems. The relative fraction of tidal features will thus be given by the product of the relative lifetimes of the tidal features with the relative number densities of such galaxies. 
Hence, for a $2.7\times{}10^{11}$ $M_\odot$ elliptical galaxy the relative fraction of tidal tails generated by encounters with disks, relative to encounters with other early types, is found to be $>1.4$ for a 1:4 mass ratio, and $>4$ for a 1:10 mass ratio. Therefore, a lower limit is that more than $60\%$ (mass ratio 1:4) or $80\%$ (mass ratio 1:10) of the visible tidal features will stem from a merger with a disk-dominated satellite. 

For a fixed mass ratio, decreasing the mass of the main elliptical galaxy (i) enhances relative lifetimes, since only the strongest features remain visible; and (ii) increases the fraction of blue versus red galaxies. Both effects tend to increase the relative fraction of tidal features.
We can estimate the relative fraction $f_n(R)$ of tidal features after integrating over all possible masses of the 
elliptical host galaxy for a fixed mass ratio $R<1$. We will assume that the relative lifetime $f_t(R)$ only depends on $R$ and not 
on the mass  itself.  The relative fraction of tidal features $f(R)$ is, in this simple approximation,  given by:  $f(R)=f_n(R)f_t(R)$. In particular,
\[
f(R) = \frac{f_t(R)}{\int{}n_E(m)dm} \int{}n_E(m)\frac{n_D(mR)}{n_E(mR)}dm
\]
where $n_E$ and $n_D$ denote the number density of early-type and disk galaxies, respectively. The integrals are well-defined and need only be evaluated near the characteristic mass $M^*$ of the early-type galaxy population (e.g. the integration between $10^9$ and $10^{12}$ $M_\odot$ is accurate up to a few percent for all relevant mass ratios).
Fig.~\ref{fig:numberDens}b
shows $f_n(R)$ calculated from the luminosity functions of~\citet{2006MNRAS.368..414D} after converting them into mass functions as described above. We obtain $f_n(1:4)=2.03$ and $f_n(1:10)=2.96$.
Hence by integrating over all plausible masses of the elliptical galaxy host we estimate that the relative tidal fractions are $>2.8$ for a mass ratio of 1:4 and $>12$ for a mass ratio of 1:10.

\section{Discussion and Conclusion}
\label{sect:discussion}

We have found major differences between the elliptical-elliptical and elliptical-disk mergers in both morphology of tidal tails and their timescales of observability. Only mergers that involve a dynamically cold disk produce strong tidal features with loops and arms resembling those observed around bright ellipticals~\citep{2005AJ....130.2647V}; these features last much longer (1-2 Gyr) than the smooth features produced in hot-hot mergers.

Dynamical friction and tidal stripping determine the specifics of the merger evolution. For a given satellite galaxy type, either disk or elliptical, the tidal tails generated by less massive satellites extend to larger distances from the center of the accreting elliptical, as the smaller the satellite, the longer it needs to sink to the bottom of the gravitational potential. However, the smaller the satellite mass, the weaker the tidal debris, since less mass is available to produce the tidal features. We find that satellite masses of order one fourth of the mass of the parent elliptical exhibit the most long-lived tidal features.

Since mergers with disk-dominated, relatively low mass companions produce the strongest and longest-lasting tidal streams, such mergers are expected to be the typical events that are revealed in observations of real galaxies. In first approximation, the probability that the main elliptical merges with a companion of a given type will be proportional to the number density of galaxies of that type. The ratio of observable tidal features that are expected from mergers involving a disk galaxy, relative to mergers that involve another elliptical, can thus be approximated by the product of the ratio between the respective tidal lifetimes and the ratio of the respective number densities. 
Using published luminosity functions for early-type and disk galaxies~\citep{2006MNRAS.368..414D} and typical mass-to-light ratios~\citep{2001ApJ...550..212B}, we estimate that, at the mass scale of the host galaxy and  for a 1:4 merger, $>60\%$ of the observed tidal features stem from an encounter with a disk-dominated system; the fraction increases up to $>80\%$ for 1:10 mergers. We thus expect that the merger fraction that is estimated from observed tidal features around bright ellipticals reflects primarily the fraction of spheroids that underwent (relatively minor) accretion events of dynamically cold, disk galaxies over the past several billion years, rather than very recent major mergers between elliptical galaxies.

Our simulations show that the observable tidal debris mostly originates from regions of the disk corresponding to about 2-5 exponential scale lengths. 
Using observed photometric profiles of spiral~\citep{1994AaAS..106..451D} and elliptical~\citep{1990AJ....100.1091P} galaxies and stellar synthesis models~\citep{2003MNRAS.344.1000B}, we find that the $B-R$ color of the tidal features is rather similar to the color of the main accreting elliptical at several half-light radii from its center, where the tidal tails become visible. We estimate that both the average $B-R$ color difference between tidal debris and main elliptical's stars, and its standard deviation, are only of order $\sim{}$ 0.3 mag. The tidal tails produced by the accretion of disk galaxies are thus ``red'' in $B-R$ color.

Disk satellites responsible for the observed tidal features would likely host a gas component, which is not included in our $N$-body simulations.  Such a gas component might cause a burst of star formation in the early phases of the interaction~\citep{1991ApJ...370L..65B}. Simulations show that tidally triggered stellar bars can develop soon after the first pericenter passage and drive substantial gas inflow within the central $\sim$1 kpc of the satellite galaxy~\citep{1996ApJ...471..115B,2005ApJ...623L..67K}, where the gas is quickly exhausted in a burst of star formation.  
Star formation is also expected in the tidal tails at these early stages of the merger. Ram pressure stripping in the hot halo of  the massive elliptical will however remove the reservoir of diffuse gas and thus quench star formation at later times~\citep{2008ApJ...672L.103K}. Ram pressure becomes increasingly efficient as tidal shocks decrease the potential well of the satellite at each subsequent pericenter and is expected to remove most if not all the gas that has not plunged into its central region~\citep{2007Natur.445..738M}. This is in agreement with the absence of a  young stellar component in -- and the red colors of -- the observed tidal features. We predict, however, that very faint and diffuse gaseous tidal tails and plumes should be detectable around elliptical galaxies with red stellar tidal signatures.

After the merger the newly formed stars in a disk-dominated progenitor might affect the overall $B-R$ color of the remnant. 
Making a sensible guess for the parameters of the central starburst, an $e$-folding time of 100 Myr and an initial star formation rate of 20 $M_\odot$yr$^{-1}$, we find  that within 500 Myr the $B-R$ color of the remnant will closely resemble the original $B-R$ color of the elliptical merger progenitor to less than 0.1 mag (Fig.~\ref{fig:RedSelection}). In addition, a nonnegligible fraction of nearby elliptical galaxies are found to host a few, up to a few tens, percent of stellar mass in young stars~\citep{2000AJ....120..165T}.

The existence of red tidal features around the most massive elliptical galaxies can thus be well explained with minor accretion events of disk-dominated satellite galaxies. Consequently, our work supports an early ($z>1$) mass assembly epoch for the most massive galaxies in the Universe with a later accretion of smaller mass units that do not contribute significantly to the total mass. Future observations will have to clarify the relative importance of dry and wet mergers at such early epochs in the formation of the bright elliptical galaxy population.

\acknowledgments
{\small
R.F. acknowledges founding by the Swiss National Science Foundation. The simulations have been carried out on the 
Gonzales cluster hosted by the Informatikdienste at the ETH Zurich. Correspondence and requests for 
materials should be addressed to the first author.}


\bibliography{feldmann}

\begin{thebibliography}{58}
\expandafter\ifx\csname natexlab\endcsname\relax\def\natexlab#1{#1}\fi

\bibitem[{{Baldry} {et~al.}(2004){Baldry}, {Glazebrook}, {Brinkmann},
  {Ivezi{\'c}}, {Lupton}, {Nichol}, \& {Szalay}}]{2004ApJ...600..681B}
{Baldry}, I.~K., {Glazebrook}, K., {Brinkmann}, J., {Ivezi{\'c}}, {\v Z}.,
  {Lupton}, R.~H., {Nichol}, R.~C., \& {Szalay}, A.~S. 2004, \apj, 600, 681

\bibitem[{{Barnes} \& {Hernquist}(1996)}]{1996ApJ...471..115B}
{Barnes}, J.~E., \& {Hernquist}, L. 1996, \apj, 471, 115

\bibitem[{{Barnes} \& {Hernquist}(1991)}]{1991ApJ...370L..65B}
{Barnes}, J.~E., \& {Hernquist}, L.~E. 1991, \apjl, 370, L65

\bibitem[{{Bell} \& {de Jong}(2000)}]{2000MNRAS.312..497B}
{Bell}, E.~F., \& {de Jong}, R.~S. 2000, \mnras, 312, 497

\bibitem[{{Bell} \& {de Jong}(2001)}]{2001ApJ...550..212B}
---. 2001, \apj, 550, 212

\bibitem[{{Bell} {et~al.}(2003){Bell}, {McIntosh}, {Katz}, \&
  {Weinberg}}]{2003ApJS..149..289B}
{Bell}, E.~F., {McIntosh}, D.~H., {Katz}, N., \& {Weinberg}, M.~D. 2003, \apjs,
  149, 289

\bibitem[{{Bell} {et~al.}(2006){Bell}, {Naab}, {McIntosh}, {Somerville},
  {Caldwell}, {Barden}, {Wolf}, {Rix}, {Beckwith}, {Borch}, {H{\"a}ussler},
  {Heymans}, {Jahnke}, {Jogee}, {Koposov}, {Meisenheimer}, {Peng}, {Sanchez},
  \& {Wisotzki}}]{2006ApJ...640..241B}
{Bell}, E.~F., {Naab}, T., {McIntosh}, D.~H., {Somerville}, R.~S., {Caldwell},
  J.~A.~R., {Barden}, M., {Wolf}, C., {Rix}, H.-W., {Beckwith}, S.~V., {Borch},
  A., {H{\"a}ussler}, B., {Heymans}, C., {Jahnke}, K., {Jogee}, S., {Koposov},
  S., {Meisenheimer}, K., {Peng}, C.~Y., {Sanchez}, S.~F., \& {Wisotzki}, L.
  2006, \apj, 640, 241

\bibitem[{{Bell} {et~al.}(2004){Bell}, {Wolf}, {Meisenheimer}, {Rix}, {Borch},
  {Dye}, {Kleinheinrich}, {Wisotzki}, \& {McIntosh}}]{2004ApJ...608..752B}
{Bell}, E.~F., {Wolf}, C., {Meisenheimer}, K., {Rix}, H.-W., {Borch}, A.,
  {Dye}, S., {Kleinheinrich}, M., {Wisotzki}, L., \& {McIntosh}, D.~H. 2004,
  \apj, 608, 752

\bibitem[{{Bender} {et~al.}(1992){Bender}, {Burstein}, \&
  {Faber}}]{1992ApJ...399..462B}
{Bender}, R., {Burstein}, D., \& {Faber}, S.~M. 1992, \apj, 399, 462

\bibitem[{{Blanton} {et~al.}(2001){Blanton}, {Dalcanton}, {Eisenstein},
  {Loveday}, {Strauss}, {SubbaRao}, {Weinberg}, {Anderson}, {Annis}, {Bahcall},
  {Bernardi}, {Brinkmann}, {Brunner}, {Burles}, {Carey}, {Castander},
  {Connolly}, {Csabai}, {Doi}, {Finkbeiner}, {Friedman}, {Frieman}, {Fukugita},
  {Gunn}, {Hennessy}, {Hindsley}, {Hogg}, {Ichikawa}, {Ivezi{\'c}}, {Kent},
  {Knapp}, {Lamb}, {Leger}, {Long}, {Lupton}, {McKay}, {Meiksin}, {Merelli},
  {Munn}, {Narayanan}, {Newcomb}, {Nichol}, {Okamura}, {Owen}, {Pier}, {Pope},
  {Postman}, {Quinn}, {Rockosi}, {Schlegel}, {Schneider}, {Shimasaku},
  {Siegmund}, {Smee}, {Snir}, {Stoughton}, {Stubbs}, {Szalay}, {Szokoly},
  {Thakar}, {Tremonti}, {Tucker}, {Uomoto}, {Vanden Berk}, {Vogeley},
  {Waddell}, {Yanny}, {Yasuda}, \& {York}}]{2001AJ....121.2358B}
{Blanton}, M.~R., {Dalcanton}, J., {Eisenstein}, D., {Loveday}, J., {Strauss},
  M.~A., {SubbaRao}, M., {Weinberg}, D.~H., {Anderson}, Jr., J.~E., {Annis},
  J., {Bahcall}, N.~A., {Bernardi}, M., {Brinkmann}, J., {Brunner}, R.~J.,
  {Burles}, S., {Carey}, L., {Castander}, F.~J., {Connolly}, A.~J., {Csabai},
  I., {Doi}, M., {Finkbeiner}, D., {Friedman}, S., {Frieman}, J.~A.,
  {Fukugita}, M., {Gunn}, J.~E., {Hennessy}, G.~S., {Hindsley}, R.~B., {Hogg},
  D.~W., {Ichikawa}, T., {Ivezi{\'c}}, {\v Z}., {Kent}, S., {Knapp}, G.~R.,
  {Lamb}, D.~Q., {Leger}, R.~F., {Long}, D.~C., {Lupton}, R.~H., {McKay},
  T.~A., {Meiksin}, A., {Merelli}, A., {Munn}, J.~A., {Narayanan}, V.,
  {Newcomb}, M., {Nichol}, R.~C., {Okamura}, S., {Owen}, R., {Pier}, J.~R.,
  {Pope}, A., {Postman}, M., {Quinn}, T., {Rockosi}, C.~M., {Schlegel}, D.~J.,
  {Schneider}, D.~P., {Shimasaku}, K., {Siegmund}, W.~A., {Smee}, S., {Snir},
  Y., {Stoughton}, C., {Stubbs}, C., {Szalay}, A.~S., {Szokoly}, G.~P.,
  {Thakar}, A.~R., {Tremonti}, C., {Tucker}, D.~L., {Uomoto}, A., {Vanden
  Berk}, D., {Vogeley}, M.~S., {Waddell}, P., {Yanny}, B., {Yasuda}, N., \&
  {York}, D.~G. 2001, \aj, 121, 2358

\bibitem[{{Bruzual} \& {Charlot}(2003)}]{2003MNRAS.344.1000B}
{Bruzual}, G., \& {Charlot}, S. 2003, \mnras, 344, 1000

\bibitem[{{Chabrier}(2003)}]{2003PASP..115..763C}
{Chabrier}, G. 2003, \pasp, 115, 763

\bibitem[{{Cimatti} {et~al.}(2006){Cimatti}, {Daddi}, \&
  {Renzini}}]{2006AaA...453L..29C}
{Cimatti}, A., {Daddi}, E., \& {Renzini}, A. 2006, \aap, 453, L29

\bibitem[{{Colpi} {et~al.}(1999){Colpi}, {Mayer}, \&
  {Governato}}]{1999ApJ...525..720C}
{Colpi}, M., {Mayer}, L., \& {Governato}, F. 1999, \apj, 525, 720

\bibitem[{{Combes} {et~al.}(1995){Combes}, {Rampazzo}, {Bonfanti}, {Pringniel},
  \& {Sulentic}}]{1995AaA...297...37C}
{Combes}, F., {Rampazzo}, R., {Bonfanti}, P.~P., {Pringniel}, P., \&
  {Sulentic}, J.~W. 1995, \aap, 297, 37

\bibitem[{{Davies} {et~al.}(1987){Davies}, {Burstein}, {Dressler}, {Faber},
  {Lynden-Bell}, {Terlevich}, \& {Wegner}}]{1987ApJS...64..581D}
{Davies}, R.~L., {Burstein}, D., {Dressler}, A., {Faber}, S.~M., {Lynden-Bell},
  D., {Terlevich}, R.~J., \& {Wegner}, G. 1987, \apjs, 64, 581

\bibitem[{{de Jong} \& {van der Kruit}(1994)}]{1994AaAS..106..451D}
{de Jong}, R.~S., \& {van der Kruit}, P.~C. 1994, \aaps, 106, 451

\bibitem[{{Djorgovski} \& {Davis}(1987)}]{1987ApJ...313...59D}
{Djorgovski}, S., \& {Davis}, M. 1987, \apj, 313, 59

\bibitem[{{Dressler} {et~al.}(1987){Dressler}, {Lynden-Bell}, {Burstein},
  {Davies}, {Faber}, {Terlevich}, \& {Wegner}}]{1987ApJ...313...42D}
{Dressler}, A., {Lynden-Bell}, D., {Burstein}, D., {Davies}, R.~L., {Faber},
  S.~M., {Terlevich}, R., \& {Wegner}, G. 1987, \apj, 313, 42

\bibitem[{{Driver} {et~al.}(2006){Driver}, {Allen}, {Graham}, {Cameron},
  {Liske}, {Ellis}, {Cross}, {De Propris}, {Phillipps}, \&
  {Couch}}]{2006MNRAS.368..414D}
{Driver}, S.~P., {Allen}, P.~D., {Graham}, A.~W., {Cameron}, E., {Liske}, J.,
  {Ellis}, S.~C., {Cross}, N.~J.~G., {De Propris}, R., {Phillipps}, S., \&
  {Couch}, W.~J. 2006, \mnras, 368, 414

\bibitem[{{Feldmann} {et~al.}(2006){Feldmann}, {Carollo}, {Porciani}, {Lilly},
  {Capak}, {Taniguchi}, {F{\`e}vre}, {Renzini}, {Scoville}, {Ajiki}, {Aussel},
  {Contini}, {McCracken}, {Mobasher}, {Murayama}, {Sanders}, {Sasaki},
  {Scarlata}, {Scodeggio}, {Shioya}, {Silverman}, {Takahashi}, {Thompson}, \&
  {Zamorani}}]{2006MNRAS.372..565F}
{Feldmann}, R., {Carollo}, C.~M., {Porciani}, C., {Lilly}, S.~J., {Capak}, P.,
  {Taniguchi}, Y., {F{\`e}vre}, O.~L., {Renzini}, A., {Scoville}, N., {Ajiki},
  M., {Aussel}, H., {Contini}, T., {McCracken}, H., {Mobasher}, B., {Murayama},
  T., {Sanders}, D., {Sasaki}, S., {Scarlata}, C., {Scodeggio}, M., {Shioya},
  Y., {Silverman}, J., {Takahashi}, M., {Thompson}, D., \& {Zamorani}, G. 2006,
  \mnras, 372, 565

\bibitem[{{Fukugita} {et~al.}(1998){Fukugita}, {Hogan}, \&
  {Peebles}}]{1998ApJ...503..518F}
{Fukugita}, M., {Hogan}, C.~J., \& {Peebles}, P.~J.~E. 1998, \apj, 503, 518

\bibitem[{{Ghigna} {et~al.}(1998){Ghigna}, {Moore}, {Governato}, {Lake},
  {Quinn}, \& {Stadel}}]{1998MNRAS.300..146G}
{Ghigna}, S., {Moore}, B., {Governato}, F., {Lake}, G., {Quinn}, T., \&
  {Stadel}, J. 1998, \mnras, 300, 146

\bibitem[{{Gnedin} {et~al.}(1999){Gnedin}, {Hernquist}, \&
  {Ostriker}}]{1999ApJ...514..109G}
{Gnedin}, O.~Y., {Hernquist}, L., \& {Ostriker}, J.~P. 1999, \apj, 514, 109

\bibitem[{{Hernquist}(1990)}]{1990ApJ...356..359H}
{Hernquist}, L. 1990, \apj, 356, 359

\bibitem[{{Hernquist}(1993)}]{1993ApJS...86..389H}
---. 1993, \apjs, 86, 389

\bibitem[{{Hernquist} \& {Quinn}(1988)}]{1988ApJ...331..682H}
{Hernquist}, L., \& {Quinn}, P.~J. 1988, \apj, 331, 682

\bibitem[{{Hibbard} \& {van Gorkom}(1996)}]{1996AJ....111..655H}
{Hibbard}, J.~E., \& {van Gorkom}, J.~H. 1996, \aj, 111, 655

\bibitem[{{Kawata} \& {Mulchaey}(2008)}]{2008ApJ...672L.103K}
{Kawata}, D., \& {Mulchaey}, J.~S. 2008, \apjl, 672, L103

\bibitem[{{Kawata} {et~al.}(2006){Kawata}, {Mulchaey}, {Gibson}, \&
  {S{\'a}nchez-Bl{\'a}zquez}}]{2006ApJ...648..969K}
{Kawata}, D., {Mulchaey}, J.~S., {Gibson}, B.~K., \&
  {S{\'a}nchez-Bl{\'a}zquez}, P. 2006, \apj, 648, 969

\bibitem[{{Kazantzidis} {et~al.}(2005){Kazantzidis}, {Mayer}, {Colpi}, {Madau},
  {Debattista}, {Wadsley}, {Stadel}, {Quinn}, \& {Moore}}]{2005ApJ...623L..67K}
{Kazantzidis}, S., {Mayer}, L., {Colpi}, M., {Madau}, P., {Debattista}, V.~P.,
  {Wadsley}, J., {Stadel}, J., {Quinn}, T., \& {Moore}, B. 2005, \apjl, 623,
  L67

\bibitem[{{Khochfar} \& {Burkert}(2003)}]{2003ApJ...597L.117K}
{Khochfar}, S., \& {Burkert}, A. 2003, \apjl, 597, L117

\bibitem[{{Khochfar} \& {Burkert}(2005)}]{2005MNRAS.359.1379K}
---. 2005, \mnras, 359, 1379

\bibitem[{{Khochfar} \& {Burkert}(2006)}]{2006A&A...445..403K}
---. 2006, \aap, 445, 403

\bibitem[{{Kodama} {et~al.}(1998){Kodama}, {Arimoto}, {Barger}, \&
  {Arag'on-Salamanca}}]{1998A&A...334...99K}
{Kodama}, T., {Arimoto}, N., {Barger}, A.~J., \& {Arag'on-Salamanca}, A. 1998,
  \aap, 334, 99

\bibitem[{{Lilly} {et~al.}(1998){Lilly}, {Schade}, {Ellis}, {Le Fevre},
  {Brinchmann}, {Tresse}, {Abraham}, {Hammer}, {Crampton}, {Colless},
  {Glazebrook}, {Mallen-Ornelas}, \& {Broadhurst}}]{1998ApJ...500...75L}
{Lilly}, S., {Schade}, D., {Ellis}, R., {Le Fevre}, O., {Brinchmann}, J.,
  {Tresse}, L., {Abraham}, R., {Hammer}, F., {Crampton}, D., {Colless}, M.,
  {Glazebrook}, K., {Mallen-Ornelas}, G., \& {Broadhurst}, T. 1998, \apj, 500,
  75

\bibitem[{{Maraston}(2005)}]{2005MNRAS.362..799M}
{Maraston}, C. 2005, \mnras, 362, 799

\bibitem[{{Mayer} {et~al.}(2007){Mayer}, {Kazantzidis}, {Mastropietro}, \&
  {Wadsley}}]{2007Natur.445..738M}
{Mayer}, L., {Kazantzidis}, S., {Mastropietro}, C., \& {Wadsley}, J. 2007,
  \nat, 445, 738

\bibitem[{{Mayer} {et~al.}(2006){Mayer}, {Mastropietro}, {Wadsley}, {Stadel},
  \& {Moore}}]{2006MNRAS.369.1021M}
{Mayer}, L., {Mastropietro}, C., {Wadsley}, J., {Stadel}, J., \& {Moore}, B.
  2006, \mnras, 369, 1021

\bibitem[{{Mayer} {et~al.}(2002){Mayer}, {Moore}, {Quinn}, {Governato}, \&
  {Stadel}}]{2002MNRAS.336..119M}
{Mayer}, L., {Moore}, B., {Quinn}, T., {Governato}, F., \& {Stadel}, J. 2002,
  \mnras, 336, 119

\bibitem[{{Michard}(1999)}]{1999A&AS..137..245M}
{Michard}, R. 1999, \aaps, 137, 245

\bibitem[{{Mo} {et~al.}(1998){Mo}, {Mao}, \& {White}}]{1998MNRAS.295..319M}
{Mo}, H.~J., {Mao}, S., \& {White}, S.~D.~M. 1998, \mnras, 295, 319

\bibitem[{{Moore} {et~al.}(1996){Moore}, {Katz}, \&
  {Lake}}]{1996ApJ...457..455M}
{Moore}, B., {Katz}, N., \& {Lake}, G. 1996, \apj, 457, 455

\bibitem[{{Moore} {et~al.}(2004){Moore}, {Kazantzidis}, {Diemand}, \&
  {Stadel}}]{2004MNRAS.354..522M}
{Moore}, B., {Kazantzidis}, S., {Diemand}, J., \& {Stadel}, J. 2004, \mnras,
  354, 522

\bibitem[{{Naab} {et~al.}(2007){Naab}, {Johansson}, {Ostriker}, \&
  {Efstathiou}}]{2007ApJ...658..710N}
{Naab}, T., {Johansson}, P.~H., {Ostriker}, J.~P., \& {Efstathiou}, G. 2007,
  \apj, 658, 710

\bibitem[{{Naab} {et~al.}(2006){Naab}, {Khochfar}, \&
  {Burkert}}]{2006ApJ...636L..81N}
{Naab}, T., {Khochfar}, S., \& {Burkert}, A. 2006, \apjl, 636, L81

\bibitem[{{Navarro} {et~al.}(1997){Navarro}, {Frenk}, \&
  {White}}]{1997ApJ...490..493N}
{Navarro}, J.~F., {Frenk}, C.~S., \& {White}, S.~D.~M. 1997, \apj, 490, 493

\bibitem[{{Peletier} {et~al.}(1990){Peletier}, {Davies}, {Illingworth},
  {Davis}, \& {Cawson}}]{1990AJ....100.1091P}
{Peletier}, R.~F., {Davies}, R.~L., {Illingworth}, G.~D., {Davis}, L.~E., \&
  {Cawson}, M. 1990, \aj, 100, 1091

\bibitem[{{Renzini}(2006)}]{2006ARAaA..44..141R}
{Renzini}, A. 2006, \araa, 44, 141

\bibitem[{{Scarlata} {et~al.}(2007){Scarlata}, {Carollo}, {Lilly}, {Feldmann},
  {Kampczyk}, {Renzini}, {Cimatti}, {Halliday}, {Daddi}, {Sargent},
  {Koekemoer}, {Scoville}, {Kneib}, {Leauthaud}, {Massey}, {Rhodes}, {Tasca},
  {Capak}, {McCracken}, {Mobasher}, {Taniguchi}, {Thompson}, {Ajiki}, {Aussel},
  {Murayama}, {Sanders}, {Sasaki}, {Shioya}, \&
  {Takahashi}}]{2007ApJS..172..494S}
{Scarlata}, C., {Carollo}, C.~M., {Lilly}, S.~J., {Feldmann}, R., {Kampczyk},
  P., {Renzini}, A., {Cimatti}, A., {Halliday}, C., {Daddi}, E., {Sargent},
  M.~T., {Koekemoer}, A., {Scoville}, N., {Kneib}, J.-P., {Leauthaud}, A.,
  {Massey}, R., {Rhodes}, J., {Tasca}, L., {Capak}, P., {McCracken}, H.~J.,
  {Mobasher}, B., {Taniguchi}, Y., {Thompson}, D., {Ajiki}, M., {Aussel}, H.,
  {Murayama}, T., {Sanders}, D.~B., {Sasaki}, S., {Shioya}, Y., \& {Takahashi},
  M. 2007, \apjs, 172, 494

\bibitem[{{Schombert} {et~al.}(1990){Schombert}, {Wallin}, \&
  {Struck-Marcell}}]{1990AJ.....99..497S}
{Schombert}, J.~M., {Wallin}, J.~F., \& {Struck-Marcell}, C. 1990, \aj, 99, 497

\bibitem[{{Stadel}(2001)}]{2001PhDT........21S}
{Stadel}, J.~G. 2001, Ph.D.~Thesis, University of Washington

\bibitem[{{Taffoni} {et~al.}(2003){Taffoni}, {Mayer}, {Colpi}, \&
  {Governato}}]{2003MNRAS.341..434T}
{Taffoni}, G., {Mayer}, L., {Colpi}, M., \& {Governato}, F. 2003, \mnras, 341,
  434

\bibitem[{{Tamm} {et~al.}(2007){Tamm}, {Tempel}, \&
  {Tenjes}}]{2007arXiv0707.4375T}
{Tamm}, A., {Tempel}, E., \& {Tenjes}, P. 2007, ArXiv e-prints, 0707.4375

\bibitem[{{Trager} {et~al.}(2000){Trager}, {Faber}, {Worthey}, \&
  {Gonz{\'a}lez}}]{2000AJ....120..165T}
{Trager}, S.~C., {Faber}, S.~M., {Worthey}, G., \& {Gonz{\'a}lez}, J.~J. 2000,
  \aj, 120, 165

\bibitem[{{van Dokkum}(2005)}]{2005AJ....130.2647V}
{van Dokkum}, P.~G. 2005, \aj, 130, 2647

\bibitem[{{Wadsley} {et~al.}(2004){Wadsley}, {Stadel}, \&
  {Quinn}}]{2004NewA....9..137W}
{Wadsley}, J.~W., {Stadel}, J., \& {Quinn}, T. 2004, New Astronomy, 9, 137

\bibitem[{{Walterbos} \& {Kennicutt}(1988)}]{1988A&A...198...61W}
{Walterbos}, R.~A.~M., \& {Kennicutt}, Jr., R.~C. 1988, \aap, 198, 61

\end{thebibliography}

\end{document}